\documentclass[prb, twocolumn, preprintnumbers, amsmath,amssymb, superscriptaddress,longbibliography,aps]{revtex4-2}

\usepackage{xr}
\usepackage{amsmath}
\usepackage{amsfonts, amssymb,amsxtra}
\usepackage[]{graphicx}
\usepackage{grffile}
\usepackage{amsfonts}
\usepackage{framed}
\usepackage{bbm}
\usepackage{braket}
\usepackage{xcolor}
\usepackage{LatexCommands}
\usepackage{chngcntr}

\counterwithout{equation}{section} 

\usepackage{bm}
\usepackage{epsfig}
\usepackage{blindtext}
\usepackage[most]{tcolorbox}
\usepackage{floatrow}
\usepackage{verbatim}
\usepackage{color}
\usepackage{diagbox}
\usepackage{outlines}

\usepackage[colorlinks=true]{hyperref}
\hypersetup{
    unicode=false,          
    pdftoolbar=true,        
    pdfmenubar=true,        
    pdffitwindow=false,     
    pdfstartview={FitH},    
    pdftitle={},    
    pdfauthor={},     
    pdfsubject={},   
    pdfcreator={},   
    pdfproducer={}, 
    pdfkeywords={} {} {}, 
    pdfnewwindow=true,      
    colorlinks=true,       
    linkcolor=magenta, 
    citecolor=blue,        
    filecolor=magenta,      
    urlcolor=blue           
}

\usepackage[normalem]{ulem}

\def\XXint#1#2#3{{\setbox0=\hbox{$#1{#2#3}{\int}$}
     \vcenter{\hbox{$#2#3$}}\kern-.5\wd0}}

\DeclareMathOperator{\tpsi}{\tilde{\psi}}

\begin{document}
\title{Emergent moments in a Hund's impurity}
\author{Victor Drouin-Touchette}
\affiliation{Center for Materials Theory, Rutgers University, Piscataway, New Jersey 08854, USA}

\author{Elio J. K{\"o}nig}
\affiliation{Center for Materials Theory, Rutgers University, Piscataway, New Jersey 08854, USA}
\affiliation{Max-Planck Institute for Solid State Research, 70569 Stuttgart, Germany}

\author{Yashar Komijani}
\affiliation{Center for Materials Theory, Rutgers University, Piscataway, New Jersey 08854, USA}
\affiliation{Department of Physics, University of Cincinnati, Cincinnati, Ohio 45221-0011, USA}

\author{Piers Coleman}
\affiliation{Center for Materials Theory, Rutgers University, Piscataway, New Jersey 08854, USA}
\affiliation{Hubbard Theory Consortium and Department of Physics, Royal Holloway, University of London, Egham, Surrey TW20 0EX, UK}

\begin{abstract}
Motivated by the relevance of Hund’s coupling in the context of
multiorbital superconductors, we revisit the problem of a multiorbital
Kondo impurity with Hund’s interaction. Using dynamical large-N
techniques, we propose an efficient approach that retains the
essential physics at play, while providing a pathway to scalable
quantum impurity solvers. We are able to follow the ground state, dynamic, and thermodynamic
properties of this system over many decades of
temperature. Our approach captures the emergence of large moments,
and follows the stretched evolution of the physics
down to their exponentially suppressed Kondo
temperature. We focus our analysis on the
intermediate finite temperature phase which presents an alternate
paramagnetic state due to the emergent moment, and discuss the
relevance of this regime to Hund's metals.
\end{abstract}

\date{\today}
\maketitle


\section{Introduction}

In multiorbital systems, the Coulomb interaction is  manifested
as a range of competing
interactions between electrons in various orbitals, known collectively
as ``Hund's interactions''.  The key component of the Hund's
interactions, the direct ferromagnetic coupling between spins, gives
rise to the well-known Hund's rules, whereby
the total spin of the multiorbital atom is maximized
- but when immersed inside a metal, the Hund's interactions become 
an important driver of complex electronic
states. These effects are particularly notable 
in the iron based family of pnictide and chalcogenide high-temperature
superconductors, and in various ruthenate metals, now known collectively 
 as ``Hund's
metals''~\cite{haule2009coherence,yin2012fractional,fanfarillo2015electronic}. Hund's metals
delay the formation of a Fermi liquid until remarkably low temperatures. The
intermediate energy scales are characterized as spin-frozen~\cite{werner2008spin}, with slowly fluctuating
magnetic moments~\cite{watzenbock2020characteristic}.

One of many challenges posed by these materials, is the task of 
developing lightweight methods to model 
the local dynamics of  Hund's metals. 
Here we approach this problem from the perspective of
multi-orbital Kondo
models~\cite{nevidomskyy2009kondo,HuangWerner2014,aron2015analytic,khajetooriansWiebe2015,horvat2019non}
which capture both the ``Hundness'' and
metallicity of Hund's metals. 
We present here a lightweight method for describing the local 
physics of a Hund's coupled quantum impurity. 

Our study, motivated by the iron-based superconductors, 
\cite{si2016high, georges2013strong} considers the case of an iron atom
in a tetrahedral environent, interacting with a conduction sea. 
We investigate the case where iron is in its
Fe${}^{1+}$ state, with one electron occupying each of the three orbitals of the
$t_{2g}$ environment. A more detailed  model of the local physics at
the iron atom including valence fluctuations, which would allow
one to tune towards to more physical Fe${}^{2+}$ configuration, 
will be part of forthcoming work.  It has been shown recently that the interplay of
these elements and spin-orbit coupling~\cite{Chen2020} at the iron sites promotes a triplet resonating valence
bond state in the $t_{2g}$ orbital triad that may then escape into the
conduction sea, bringing about a fully gapped superconducting state
\cite{coleman2020triplet}. In light of this work, we here present the
first steps to tackle the Hund coupled impurity with a method that is
versatile enough to eventually bridge to more physical situations.

\begin{figure}[t]
    \centering
    \includegraphics[width=\linewidth]{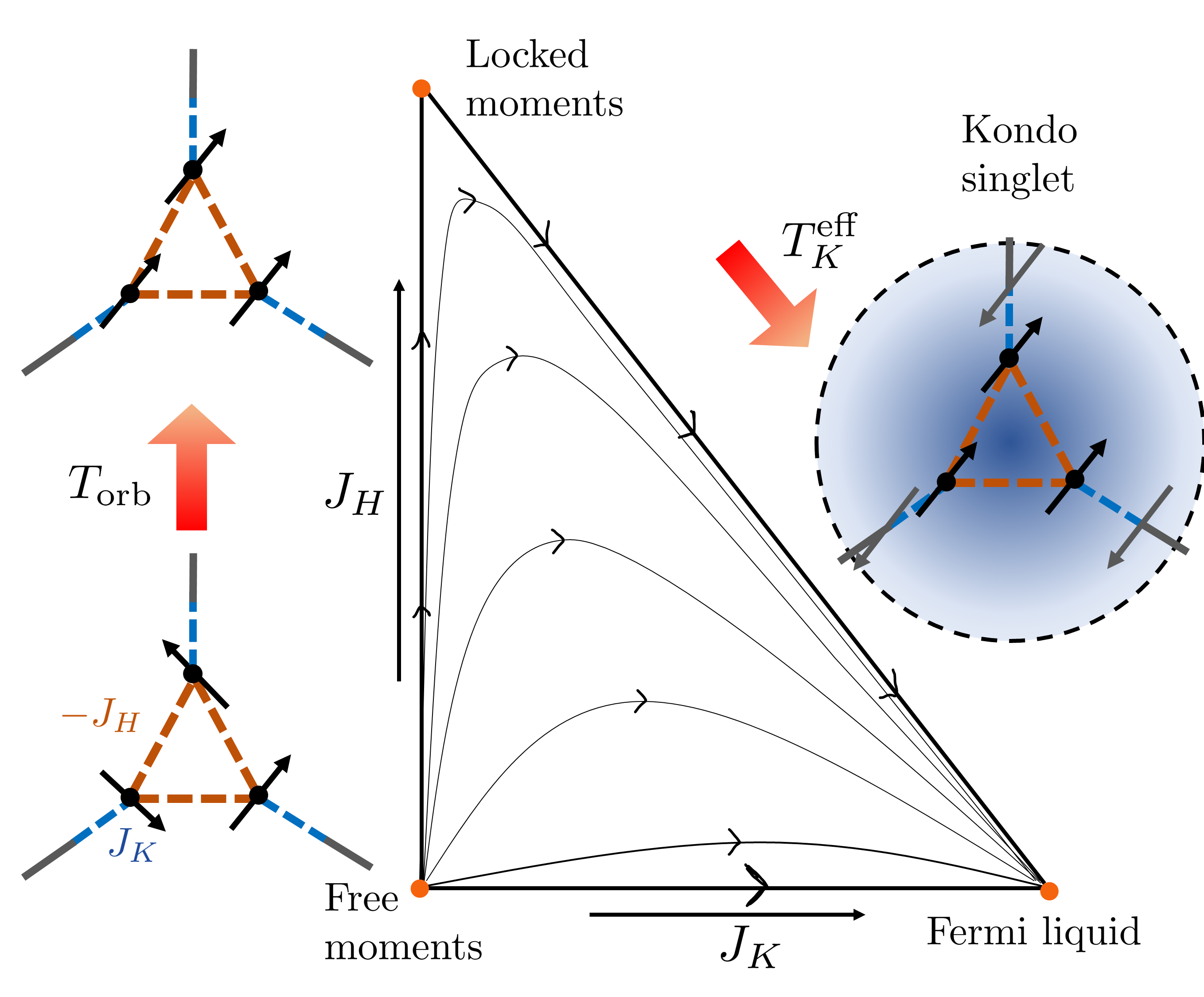}
    \caption{
    Schematic of the different physical regimes employing a conjectured renormalization group (RG)  diagram~\cite{nevidomskyy2009kondo} (arrows signifying the flow towards reduced temperature). At high temperatures, the system is in the free moment regime (near the lower left fixed point and inset). In the decoupled case, $J_H = 0$, the system quickly reaches the Nozières Fermi liquid within a relatively short RG time corresponding to $T_{K}^0 \sim D e^{- 1/\rho J_K}$. If however $J_H \gg T_K^0$, upon reduction of temperature below $T_{\rm orb} \sim J_H$, the spins of the three orbitals first align ferromagnetically (flow towards the ``locked moments'' fixed point, see top left inset). Subsequently,  at low temperatures $T < T_{K}^{\rm eff}$, the RG flow leads to the Fermi liquid state as the conduction sea fully screens the large moment (bottom right fixed point and inset).
    }
    \label{fig:figure_1}
\end{figure}

In the Hund-Kondo model, the presence of the Hund's coupling acts as a new energy scale below which a larger local moment emerges out of the impurity. This nearest-orbital interaction $-J_H$ leads to ferromagnetic alignment of the local moments between different orbitals. Furthermore, each orbital is individually coupled to an autonomous electron bath via an antiferromagnetic Kondo coupling $J_K$, recreating the local embedding of the impurity in a conduction sea, as shown through the schematics of Fig.~\ref{fig:figure_1}. The Hamiltonian is then 

\begin{equation}
\begin{split}
H_{HK} &= H_c  + \sum_{m=1}^{3} \big( J_K \vec{S}_m \cdot \vec{\sigma}_m  - J_H {\vec{S}}_m \cdot \vec{S}_{m+1} \big) \; , \\
H_c &= \sum_{m=1}^{3} \sum_{\kappa} \epsilon_{\kappa} c^\dagger_{\kappa,m}c_{\kappa,m}\; , \end{split}
\label{hk_ham}
\end{equation}

\noindent with $H_c$ the conduction-electron specific Hamiltonian, and where $\vec{S}_m$ is the spin on the $m$-th orbital (where $m$ is defined modulo $3$, i.e. $\vec S_4 \equiv \vec S_1$). The conduction electrons are coupled to the local moments $\vec{S}_m$ through their spin density $\vec{\sigma}_m \equiv c^{\dagger}_{m \alpha} \vec{\sigma}_{\alpha \beta}c_{m \beta}$  with $\alpha, \beta = \up, \down$. The conduction electron creation operator at the orbital site is given by $c^{\dagger}_{m \alpha} = \sum_{\kappa} c^{\dagger}_{\kappa m \alpha}$, where $\kappa$ is the electron momentum such that the electron's spectrum is $\epsilon_{\kappa}$. The same model with antiferromagnetic $J_H < 0$ was recently studied in ref.~\cite{konig2020frustrated} as a toy model for deconfinement. 

The structure of the paper is as follows. In section~\ref{sec1}, we review the current literature concerning the Hund-Kondo model. We present a heuristic interpretation of the different phases as well as the phase diagram obtained using dynamical large-N formalism. The analytical methods used are briefly covered in section~\ref{sec2}. A thorough analysis of the thermodynamical observables and spectral features obtained from the large-N treatment is presented in section~\ref{sec3}, as well as a comparison to expected limiting values. An approximate analytical treatment of the self-consistency equations is presented in section~\ref{sec4}. This single-iteration approach confirms the general features of the phase diagram obtained through our numerical investigation. Finally, section~\ref{secEnd} summarizes our results and provides a discussion of the successes and remaining challenges of this method to tackle realistic Hund's coupled multiorbital systems.

\section{Phase diagram overview} \label{sec1}

The key aspects of the Hund-Kondo model can be pictorially summarized in a schematic renormalization group (RG) flow diagram, Fig.~\ref{fig:figure_1}. As you will see, this model presents 3 fixed points: a free moments fixed point ($J_H = J_K = 0$), a locked moments fixed point ($J_H \neq 0$, $J_K = 0$), and an attractive Fermi-liquid fixed point ($J_H = 0$, $J_K \neq 0$). Hence, the thermodynamics are inherently dictated by the tuning of $J_H$ and $T_K^0 \simeq D e^{-1/\rho J_K}$ with $\rho =1/2D$ the uniform electronic density of states for a conduction electron band with half bandwidth $D$. Two extreme limits are readily explained. For $J_H \ll T_K^0$, the spins  magnetically decouple, amounting to three copies of the standard single channel Kondo model. Hence, the ground state is a product state of three singlets between local moments on the orbitals $\vec{S}_m$ and the conduction electrons spin density $\vec{\sigma}_m$, generating the Nozières Fermi liquid~\cite{nozieres1980kondo} at temperatures below $T = T_K^0$. The coupling between adjacent spins is irrelevant near this fixed point. This amounts to the exclusively horizontal flow in Fig.~\ref{fig:figure_1}.

The opposite limit $T_K^0 \ll J_H $ and $J_H > D$ was investigated by Schrieffer in 1967 \cite{schrieffer1967kondo}, who was studying the dependence of the Kondo temperature on the impurity spin. This situation corresponds to a flow that starts at the locked moment regime, and then flows directly to the Fermi Liquid point. This is because such a large $J_H$ leads to an automatic alignment of the three local moments into a larger spin ${S}' = 3 {S}$ moment. The larger spin ${S}'$ moment now interacts with the total spin density $\vec{\sigma} = \sum_m \vec{\sigma}_m$ via an effective Kondo coupling $J_{K,\rm eff} = J_K/3$. This leads to $T_K^{\rm eff}/D = (T_K^0/D)^M$, where $M=3$ for the triad, showcasing an exponential suppression of the Kondo temperature. For systems of $M$ strongly Hund-coupled orbitals, this result predicts the suppression of the Kondo temperature through five orders of magnitude between Ti${}^{2+}$ ($M=2$, $S=1$) and Mn${}^{2+}$ ($M=5$, $S=5/2$) impurities.

The intermediate regime where $J_H$ plays a significant role but does not exceed the electronic bandwidth is the prime interest of this work. Studies using Poor Man's scaling \cite{nevidomskyy2009kondo} as well as numerical RG \cite{de2011janus} show that as temperature is brought down, the effective moment $\mu$ (that is extracted from the spin susceptibility $\chi \sim \mu^2/T$) goes from a high-temperature value $\mu_{\rm I}$ to an intermediate larger value $\mu_{\rm II} > \mu_{\rm I}$. This larger emergent moment then has the effect of drastically reducing the Kondo temperature. This model, with its rich physics and understood phases, is a prime setting for testing the ability of our approach to extract detailed thermodynamical information and dynamical correlations.

\begin{figure*}[tb]
    \centering
    \includegraphics[width=\linewidth,trim=4 4 4 4,clip]{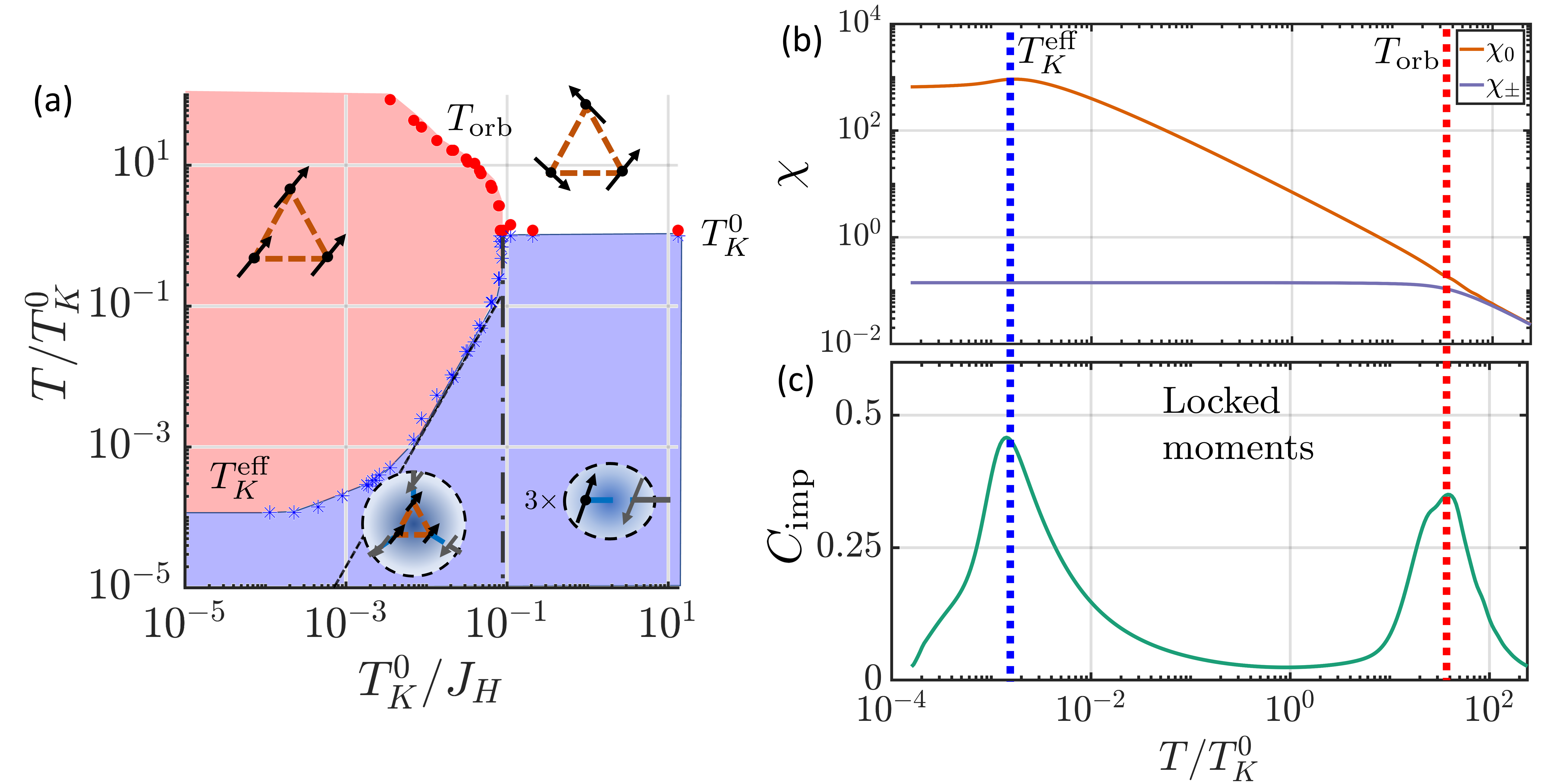}
    \caption{ (a) Phase diagram obtained by self-consistently solving equations \eqref{eq:selfcons}. Data points are extracted from local maxima of the specific heat, cf. (c), on the right, with blue for the low-temperature maxima at $T_K^{\rm eff}$ and red for the high-temperature maxima at $T_{\rm orb}$. The blue crosses (red dots) in the diagram correspond to the Fermi-Liquid (locked moment) phase boundary, respectively. The right side of the $T_K^0/J_H$ axis corresponds to the uncoupled moments, whereas the left side corresponds to the Schrieffer limit which stops the downward trend of $T_K^{\rm eff}$. The black dashed  line at intermediate $J_H$ follows the form $ T \simeq T_K^0 (T_K^0/J_H)^2$. The grey dashed-dotted line delineates the cross-over between the high and low spin limits of the Nozi\`{e}res Fermi liquid. (b) The onsite ($\chi_0$) and intersite ($\chi_{\pm}$) susceptibilities for $T_K^{0}/J_H = 0.07$.  (c) The impurity specific heat $C_{\rm imp}$ for the same parameters. Two clear peaks of the specific heat identify the boundaries $T_K^{\rm eff}$ and $T_{\rm orb}$ of the emergent moment regime, which is also seen by a plateauing of the orbital susceptibility. Sketches for each phases are added. Once the Fermi Liquid is established below $T_K^{\rm eff}$, the uniform susceptibility becomes Pauli-like. All results were obtained with the local filling of Schwinger bosons set to $q = 2S/N = 0.3$.}
        \label{fig:figure_2}
\end{figure*}

We choose to revisit this model using a large-N Schwinger boson approach that treats the ferromagnetic Hund's coupling and the antiferromagnetic Kondo coupling on the same footing. This method was successfully used to treat both ferromagnetic and antiferromagnetic spin chains embedded in a conduction bath \cite{komijani2018model, komijani2019emergent} and in more complex models \cite{wang2020quantum,shen2020strange, komijani2020isolating}. The approach's flexibility derives from its unification of the Arovas and Auerbach treatment of ferromagnetism using Schwinger bosons \cite{arovas1988functional} with the Parcollet-Georges decoupling of the Kondo problem \cite{parcollet1997transition, parcollet1998overscreened, coleman2005sum, rech2006schwinger}. It is particularly powerful in describing the dynamics of emergent excitations.

The results of our investigation are summarized in Fig.~\ref{fig:figure_2} (a). Starting on the right side of the diagram, $T_K^0/J_H$ is large and the ferromagnetic coupling between orbitals is unable to align the local moments before they are independently screened by their individual conduction electron bath at a temperature $T \sim T_{K}^0$. Below this temperature, the system settles into a low-temperature Fermi liquid phase with Pauli spin susceptibility $\chi \sim {1}/{T_K^0}$ and linear specific heat $C_v \sim {T}/{T_K^0}$. The ground state is a triplicate of independent Kondo-singlets formed from the individual local moments and their respective conduction electron seas.

As $J_H$ is increased, the system arrives at a point where it is energetically favorable for the local moments to align with one another. The thermodynamics of this intermediate regime are shown in Fig.~\ref{fig:figure_2} (b) and (c). The locking which develops at $T_{\text{orb}} \propto J_H$ is signaled by a specific heat peak as the system screens the orbital degrees of freedom. In the locked moment phase, we observe a plateau in the intersite spin susceptibility ($\chi_{\pm} \propto \langle \vec S_{m \pm 1}(\tau)\cdot \vec S_m (0) \rangle$\;), as magnetic excitations become damped. As the temperature is further decreased, the onsite susceptibility ($\chi_0$) follows a Curie temperature dependence, $\chi_0 \sim \mu_{II}^2/T$, due to the large emergent free moment. This regime spans many decades in temperature until Kondo screening inevitably occurs at a new Kondo temperature $T_K^{\rm eff}$, heralded by a large specific heat peak and a Pauli-like spin susceptibility. This effective Kondo temperature is exponentially reduced as $J_H$ is increased. The ground state is now one Kondo-singlet formed from the emergent local moment and three conduction electron seas. This is different from the ground state seen on the right; at the large-N level, the transition between those two states is first order, whereas for finite N we expect it to become a crossover.

Finally, once $J_H > D$, there can be no free local moments on the impurity, and the high-temperature limit of the problem is one of three locked local moments. Further increasing $J_H$ leads to the same high-temperature fixed point, which ends the downward reduction of the effective Kondo temperature. Instead, we find a small, fixed Kondo temperature corresponding to the aforementioned Schrieffer limit.  

We note that we can fit the exponential decrease of $T_K^{\rm eff}$ in the limit of intermediate $J_H$ by the form ${T_{K}^{\rm eff}}/{T_K^0} \sim ({T_K^0}/{J_H})^{\zeta}$, where $\zeta = M-1$ was predicted by Coleman and Nevidomskyy \cite{nevidomskyy2009kondo}. The black line in Fig.~\ref{fig:figure_2}, corresponding to $\zeta = 2$, shows the agreement of our work with this prediction. The same behavior is obtained using a single iteration of the large-N equations as an approximation (see Sec.~\ref{sec4}). In the next sections, we first introduce in detail the large-N formulation of the Hund-Kondo model, and then our detailed thermodynamical results for the intermediate regime.


\section{Dynamical large-N approach} \label{sec2}

\subsection{Schwinger boson formulation}

The dynamical large-N approach, on which we focus our attention in this paper, is taken by a fractionalized representation of the local moments by means of Schwinger bosons. Whereas a realistic system has spins of size $S$ and of $\text{SU}(2)$ symmetry, here we generalize this symmetry group to $\text{SU}(\text{N})$. Analytical solutions are controlled as $N \rightarrow \infty$. The $m$-th local moment's representation in terms of Schwinger bosons is then 

\begin{equation}
S_{m,\alpha \beta} = b^{\dagger}_{m\alpha} b_{m \beta} - \delta_{\alpha \beta} \frac{2S}{N} \; , \label{eq:sun}
\end{equation}

\noindent with the constraint on the number of bosons per orbital, $n_b(m) = 2S$. This constraint is enforced through a Lagrange multiplier $\lambda_m$. The local moment on each orbital is individually coupled to a conduction sea with $K=2S$ channels. This is an essential element of the technique, as setting the number of channels to be commensurate with spin $K = 2S$ allows the development of a perfectly screened Kondo effect. Since the second term in equation~\eqref{eq:sun} acts as an irrelevant local scattering potential, we omit it in subsequent calculations. Under this procedure, the Hamiltonian becomes 

\begin{equation}
\begin{split}
H &= H_{c} + H_{K} +H_{H} + \sum_{m=1}^{3} \lambda_m (n_{B,m} - 2S\:) \;, \\
H_{c} &=  \sum_{\kappa,m} \epsilon_{\kappa} c^\dagger_{\kappa ma\alpha}  c_{\kappa ma\alpha} \;, \\
H_{K} &= \frac{J_K}{N} \sum_m (b^{\dagger}_{m\alpha} c_{m a \alpha} ) (c^{\dagger}_{m a \beta} b_{m \beta}) \;, \\
H_{H} &= - \frac{J_H}{N} \sum_m (b^{\dagger}_{m\alpha} b_{m+1, \alpha} ) (b^{\dagger}_{m+1,\beta} b_{m\beta}) \;.
\end{split}
\end{equation}

\noindent with the assumption of an implicit summation convention over repeated indices $a$ and $\alpha$, with the indices $\alpha \in [1,\cdots,N]$ referring to the $SU(N)$ spin indices and $a \in [1,\cdots,K]$ to the channel index. In the large-N method, the coupling constants have been scaled so that the action is extensive in $N$. To maintain perfect screening at the individual orbital sites, we take $q=k$ where $2S = qN $ and $K = kN$. Carrying out a Hubbard-Stratonovich decoupling of the interaction terms leads to

\begin{equation}
\begin{split}
H_{H} \rightarrow& \sum_m [\bar{\Delta}_m (b^{\dagger}_{m+1,\alpha} b_{m \alpha}) + h.c.] + \frac{N |\Delta_m|^2}{J_H}, \\
H_{K}\rightarrow& \sum_m  [ (b^{\dagger}_{m\alpha} c_{m a \alpha})\chi_{m a} + h.c.] + \frac{N \bar{\chi}_{m,a} \chi_{m,a}}{J_K} , 
\end{split}
\end{equation}

\noindent where we have introduced two new fields. The first one $\Delta_m$ corresponds to the hopping amplitude of the Schwinger bosons, which we will call spinons. The $\chi_{m a}$ are Grassmann fields representing the charged, spinless holons that mediate the Kondo interaction in channel $a$. The full action in the large-N limit, maintaining the implicit sum over repeated indices $a$ and $\alpha$ is then 

\begin{equation}
\begin{split}
S &= \int_0^{\beta} d\tau \sum_m \Big[ \sum_{\kappa}  c^\dagger_{\kappa,ma\alpha} (\partial_{\tau} + \epsilon_{\kappa}) c_{\kappa,ma\alpha }  \\
& + b^{\dagger}_{m,\alpha} (\partial_{\tau} + \lambda_m) b_{m \alpha} + [\bar{\Delta}_m (b^{\dagger}_{m+1,\alpha} b_{m \alpha}) + h.c.]   \\
&- 2S \lambda_m + \frac{N |\Delta_m|^2}{J_H}  \\
&+ \frac{1}{\sqrt{N}}[ (b^{\dagger}_{m\alpha} c_{m a \alpha})\chi_{m a} + h.c.] + \frac{N}{J_K} \bar{\chi}_{m,a}\chi_{m,a}   \Big] \; .
\end{split} \label{eq:action}
\end{equation}

\subsection{Self-consistent solution}

\begin{figure}[tb]
    \centering
    \includegraphics[width=0.8\linewidth]{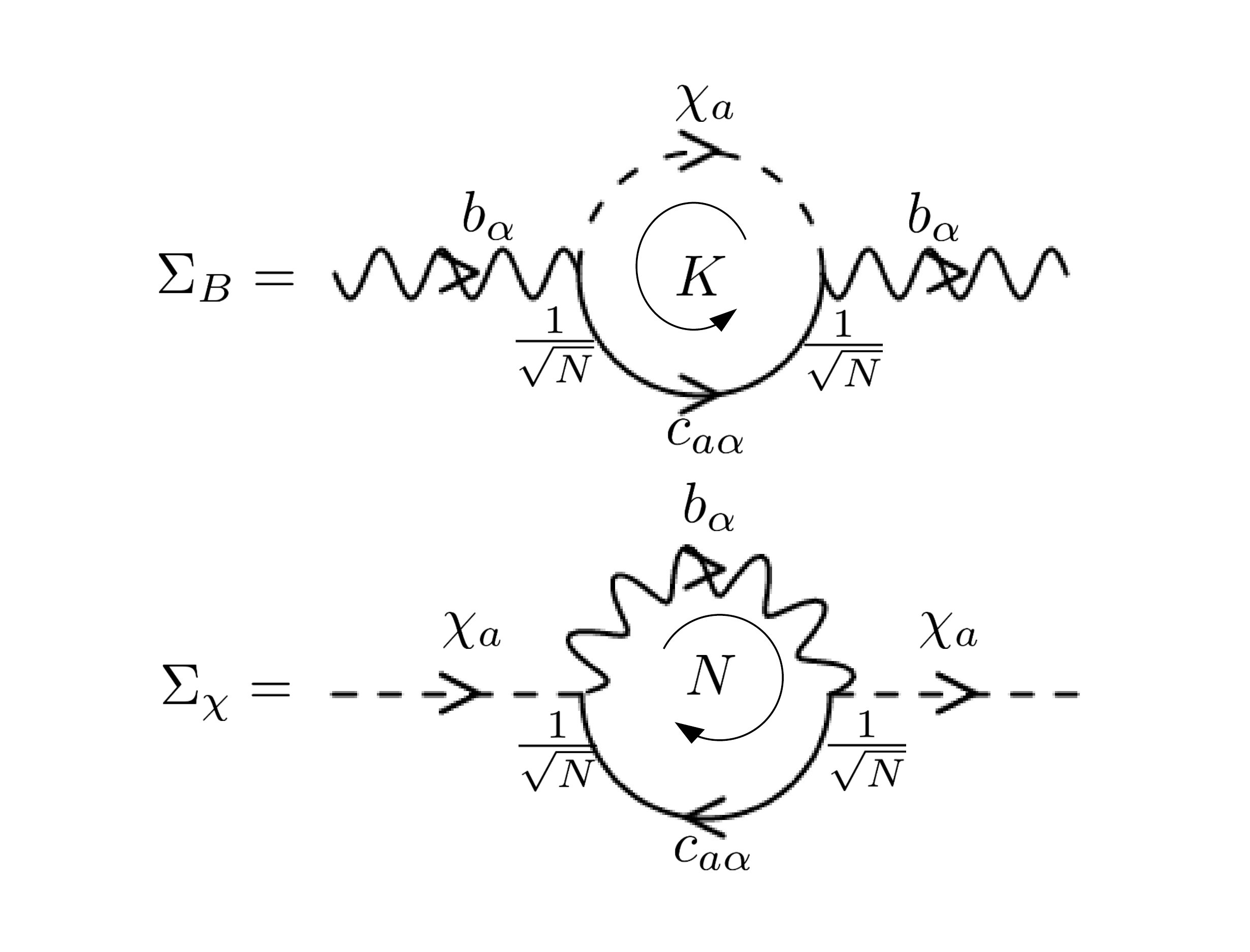}
    \caption{
    The bosonic ($\Sigma_B (\tau)$) and holonic ($\Sigma_{\chi} (\tau)$) self-energies to leading order in the $1/N$ expansion. }
    \label{fig:Feyn1}
\end{figure}

The spinon and holon have non-trivial dynamics, interacting with one another, via the $(b^{\dagger}_{m\alpha} c_{m a \alpha})\chi_{m a}$ vertex originating from the decoupling of the Kondo interaction. This causes them to influence each other's self-energy. This generates self-consistent equations for the holon and spinon self-energy, as is shown in Fig~\ref{fig:Feyn1} as well as in imaginary time in equation~\eqref{eq:selfcons}

\begin{equation}\label{eq:selfcons}
\Sigma_{\chi} (\tau) = g_c(-\tau) G_B (\tau) \; , \; \Sigma_{B} (\tau) = - k g_c(\tau) G_{\chi} (\tau) \;.
\end{equation}


These self-energy self-consistent equations are written in terms of the local propagators $G_{\chi}(\tau)$, $G_{B}(\tau)$ and $g_c(\tau)$ for the holon, spinon and conduction electrons, respectively. These two equations are of order $O(1)$ in the large-N method, since each vertex contributes $1/\sqrt{N}$ and the summation of $a$ or $\alpha$ indices in the loop leads to a further factor of $K$ or $N$, respectively. On the other hand, the conduction electron's self-energy correction is of order $O(1/N)$, which we then neglect in the large-N scheme. We discuss this further in section~\ref{secEnd}. The bare Green's function for the conduction electron $g_c(\tau) = g_{c,0} (\tau)$ is used; in the frequency domain it takes the form

\begin{equation}
g_{c,0} (z) = \int^D_{-D} d\epsilon  \frac{\rho(\epsilon)}{z-\epsilon} = -\rho \log \Big( \frac{D-z}{-D-z} \Big)  \;, \label{eq:elG}
\end{equation}
\noindent corresponding to a flat density of states with half bandwidth of $D$. The holon's Green's function takes the form 

\begin{equation}
G_{\chi} (z) = [-J_K^{-1} - \Sigma_{\chi} (z) ]^{-1} \;. \label{eq:holonG}
\end{equation}

\noindent The bare holon Green's function is devoid of any dynamics at the bare level and its dynamical features derive from its self-energy. It is also purely local (no $m$ dependence). As for the spinons, one can show (see appendix~\ref{appendix1}) that the uniform solution $\Delta_m = -\Delta$ is the lowest energy state in the absence of any Kondo coupling. Hence, we study translationally invariant solutions where, for the full temperature range studied, $\Delta_m = - \Delta$ and $\lambda_m = \lambda$. The spinons then have a discrete spectrum $\epsilon_B (p) = - 2\Delta \cos{p}$ where $p = 0, \pm 2\pi/3$. Each mode has a propagator $G_B (p, z) = [z - \epsilon_B (p) - \lambda - \Sigma_B (z)]^{-1}$. Here we used that the locality of electronic and holonic Green's functions implies that the bosonic self energy is local in orbital space. The local propagator is then

\begin{align}
&G_{B, \rm loc}(z) = \frac{1}{3}\sum_p G_B (p, z)  \label{eq:bosonG} \\
& =  \frac{1}{3} \left( \frac{2}{z - \Delta - \lambda - \Sigma_B (z)} + \frac{1}{z + 2\Delta - \lambda - \Sigma_B (z)} \right)  . \notag
\end{align}

\noindent The self-consistent equations presented in Eq.~\eqref{eq:selfcons} need to be solved while preserving the stationarity of the free energy with respect to variations of $\lambda$ and $\Delta$. This leads to the following two constraints

\begin{align}
q &= -\int^{\infty}_{-\infty} \frac{d\omega}{\pi} n_B (\omega) \Imp [G_{B, \rm loc} (\omega + i \eta)] \label{eq:qDelta} \\
\frac{\Delta}{J_H} &= \sum_{p} \int^{\infty}_{-\infty} \frac{d\omega}{\pi} n_B (\omega) \Imp [ \cos{(p)} \;  G_B (p, \omega + i \eta)] \; . \notag
\end{align}

In the numerical solution of these equations, we set $\Delta$ and $q = 2S/N$ as initial parameters. We subsequently use the constraints, Eq.~\eqref{eq:qDelta}, to determine the corresponding chemical potential $\lambda$ and the Hund's coupling $J_H$. The self-consistency equations, Eq.~\eqref{eq:selfcons}, are solved using the real frequency form of $G_B$ and $G_{\chi}$ on a logarithmic grid spanning $\omega/D \in \pm [ 10^{-5} , 10^{5}]$. All thermodynamic quantities are extracted from the real frequency form. 

Large-N treatments of ferromagnetism are notoriously plagued by 1st-order transitions. To circumvent such artefacts we add a repulsive biquadratic term $H'_m = \xi J_H (\vec{S}_m \cdot \vec{S}_{m+1})^2$. For SU(2) spins, this perturbation can be reabsorbed in the nearest neighbor interaction, but for SU(N) spins, it leads to a quartic term in the effective action and transforms the 1st order features into 2nd order transitions. Upon adding such a term, the effective $J_H^{\ast}$ is obtained as $J_H^{\ast} = J_H/(1+\xi \Delta^2/J_H^2)$, which removes the non-monotonic features \cite{komijani2018model,shen2020strange}; a value of $\xi = 10$ has been used here.


\section{Results} \label{sec3}

\subsection{Observables}

In this study, we concentrate on three observables (impurity entropy, specific heat and susceptibility), in addition to the spectral functions of the emergent spinon and holon excitations.

Notably, we obtain the entropy and the specific heat from an explicit formula derived in Refs.~\cite{coleman2005sum, rech2006schwinger} (see appendix \ref{appendix3} for a summary). The spin susceptibility $\chi(n, \tau) = \sum_{m, \alpha, \beta} \langle S_{m+n,\alpha \beta}(\tau) S_{m,\beta \alpha}(0) \rangle/3N^2$ contains an onsite part $\chi_{\rm loc}(\tau)  \equiv \chi(n =0, \tau)$ and an intersite part $\chi_{\pm}(\tau) \equiv 3\chi_{\rm loc} - \chi(n = \pm 1, \tau) $. It will be convenient to consider the Fourier transform of the susceptibility, which may be expressed \cite{komijani2018model} as

\begin{align}
&\chi (k, \omega) = \frac{1}{3} \sum_p  \int \frac{d\omega'}{2\pi} n_B (\omega') \Imp [G_B (p,\omega' +i \eta)]  \label{eq:susc} \\
& \times \left[ G_B (p - k,\omega' - \omega - i\eta)  + G_B (p+k,\omega' + \omega + i\eta) \right]  .  \notag
\end{align}

Three main static susceptibilities ($\omega \rightarrow 0$) are extracted: the zero-momentum, local and finite momentum susceptibilities. 

\begin{subequations}
\begin{align}
\chi_0 &= \chi(k=0, \omega =0) = \sum_n \chi(n, \omega = 0),\\
\chi_{\rm loc} &= \sum_k  \chi(k , \omega =0) = \chi(n = 0, \omega = 0), \\
\chi_{\pm} &= \chi(k = \pm 2\pi/3, \omega = 0) = [3\chi_{\rm loc} - \chi_{0}]/2,
\end{align}
where in terms of Schwinger bosons,
\begin{align}
\chi_{\rm loc} &=  \int \frac{d\omega}{\pi} n_B(\omega) \Imp G_{B, \rm loc}^2 (\omega + i \eta), \\
\chi_{0} &=  \frac{1}{3} \sum_p \int \frac{d\omega}{\pi} n_B(\omega) \Imp G_{B, \rm loc}^2 (p,\omega + i \eta), 
\end{align}
\end{subequations}

In the three-site impurity model, only the $k = \pm 2\pi/3$ finite momenta are present, and both momenta have the same form. In this situation, the finite momentum susceptibility corresponds to the intersite susceptibility.

Both the local and uniform susceptibility exhibit Curie-like behavior at high-temperature characteristic of a paramagnet state, as well as a Pauli-like behavior at low temperature once the Kondo-effect sets-in. Ahead of computations, it is possible to extract limiting values for the susceptibilities as well as the total impurity entropy in the large-N limit. These derivations are presented in appendix~\ref{appendix3}. Two distinct regimes of $J_H \gg T$ and $T \gg J_H$ can be tackled, and an analytic form for each regime is obtained. Both limits can be understood in terms of general functions 

\begin{align}
\tilde{S} (x) &= (1+x) \ln{(1+x)} - x \ln{x}, \\
\tilde{\chi} (x,T) &= 2x(1+x)/T,
\end{align}

\noindent with $\tilde{S}(q)$ the high-temperature entropy for a system system with $q = 2S/N$, and $\tilde{\chi} (q,T)$ the high-temperature spin susceptibility in its Curie form. The analytic limiting results are summarized in table~\ref{tableA}. This analysis reveals that the formation of the large moment for $J_H \gg T$ reduces the entropy as the system has an effective $q' = 3 q$. This also leads to a plateau in the finite-momentum susceptibility, as all temperature dependence is overshadowed by the large spinon hopping $\Delta$ present.

\renewcommand\arraystretch{2}

\begin{table}[h]
\centering
\begin{tabular}{ |c||c|c|  }
 \hline
  & $J_H \gg T$ & $T \gg J_H$\\
 \hline \hline
 $\chi_0$   &  $\tilde{\chi}(3q,T)/3$    & $\tilde{\chi}(q,T)$ \\
 \hline 
 $\chi_{\pm}$   & $\frac{2q}{\Delta}$    & $\tilde{\chi}(q,T)$\\
 \hline
 $\chi_{loc}$  &  $\frac{1}{3}[ \frac{\tilde{\chi}(3q,T)}{3} + \frac{4q}{\Delta}]$  &  $\tilde{\chi}(q,T)$ \\
 \hline
 $S_{\text{imp}}$ & $\tilde{S}(3q)$ & $3\:\tilde{S}(q)$ \\
 \hline
\end{tabular}
 \caption{The limiting values of the susceptibilities and the impurity entropy, with $q = 2S/N$ and $\beta = 1/T$ the inverse temperature. Derivation of these limits is presented in appendix~\ref{appendix3}.}
  \label{tableA}
\end{table}

\subsection{Numerical solution}

\begin{figure}[tb]
    \centering
    \includegraphics[width=\linewidth]{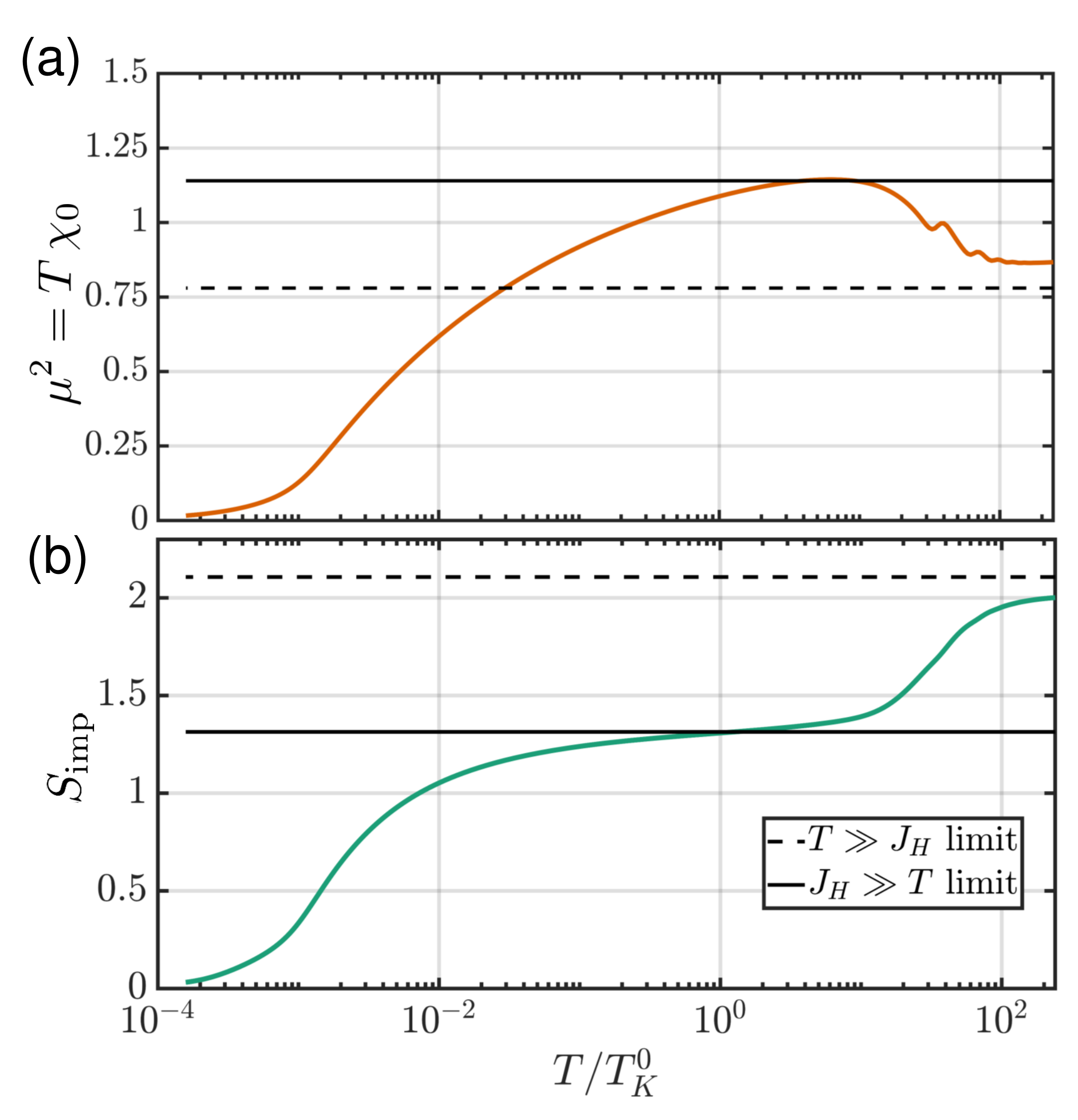}
    \caption{
        (a) The impurity moment $\mu^2 = T\chi_0$ and (b) the entropy $S_{\rm imp}$, respectively, as well as the two limits obtained analytically, as seen in Tab~\ref{tableA}. Parameters used were $q = 0.3$ and $T_K^{0}/J_H = 0.07$.}
    \label{fig:figure_4}
\end{figure}

The obtained thermodynamical quantities for a prototypical point in the intermediate regime are presented in Fig.~\ref{fig:figure_4}. From the analytical expression of the impurity entropy in the large-N limit, it is possible to directly extract the entropy $S_{\rm imp}(T)$ and specific heat $C_{\rm imp}(T)$, from its derivative. The specific heat data was used to construct our phase diagram, see Fig.~\ref{fig:figure_2}. Furthermore, we observe that when the system loses a significant proportion of its high-temperature entropy, the uniform susceptibility $\chi_0$ changes behavior. While remaining Curie-like such that $\chi_0 \sim \mu^2/T$, the moment size increases, as is seen in Fig.~\ref{fig:figure_4} (a). Dotted and filled black lines correspond to the two analytical limits for the entropy and the magnetic moment, and these are indeed reached in their respective limits. 

\begin{figure}[tb]
    \centering
    \includegraphics[width=\linewidth]{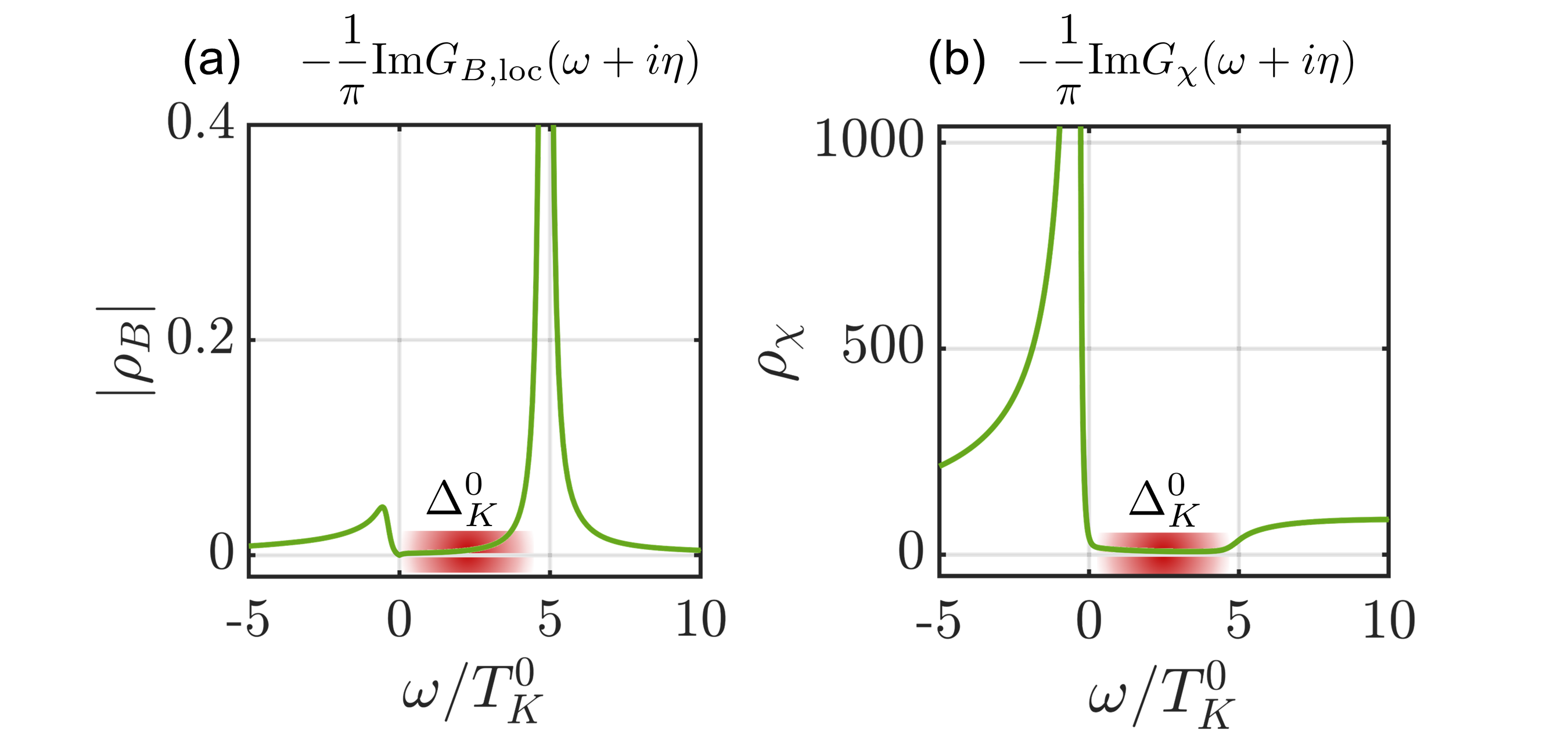}
    \caption{ Holon and spinon spectral function ((a) and (b) respectively) for $T_K^{0}/J_H = 100$ (corresponding to $\Delta/T_K^0 = 0.01$) at $T = 0.0945 T_{K}^{0}$, with a clear Kondo gap $\Delta_K^0$ at small frequencies. Parameters used were $q = 0.3$.}
    \label{fig:figure_31}
\end{figure}

Solving the self-consistent equations in the uncoupled limit ($T_K^{0}/J_H \gg 1$), we observe that changes of the Hund's coupling do not change the thermodynamics, and all curves collapse onto each other. In all cases, as the temperature is lowered, each orbital moment becomes fully screened and forms a Fermi-liquid. The Schwinger bosons and the holons both present a strong spectral gap $\Delta_K^{0} \simeq T_K^{0}$ for temperatures below the bare Kondo temperature. This is seen in Fig.~\ref{fig:figure_31} (a). The strong coupling of the spinons and the conduction electrons into a singlet, which manifests itself as a full holon phase shift of $\delta_{\chi} \equiv \Imp \ln [-G_{\chi}^{-1} (0 - i \eta)] = \pi$, is the reason for this gap. Because of the Friedel sum rule between electrons and holons, this indicates an electronic phase shift of $\delta_c = \pi/N$.

\begin{figure}[tb]
    \centering
    \includegraphics[width=\linewidth]{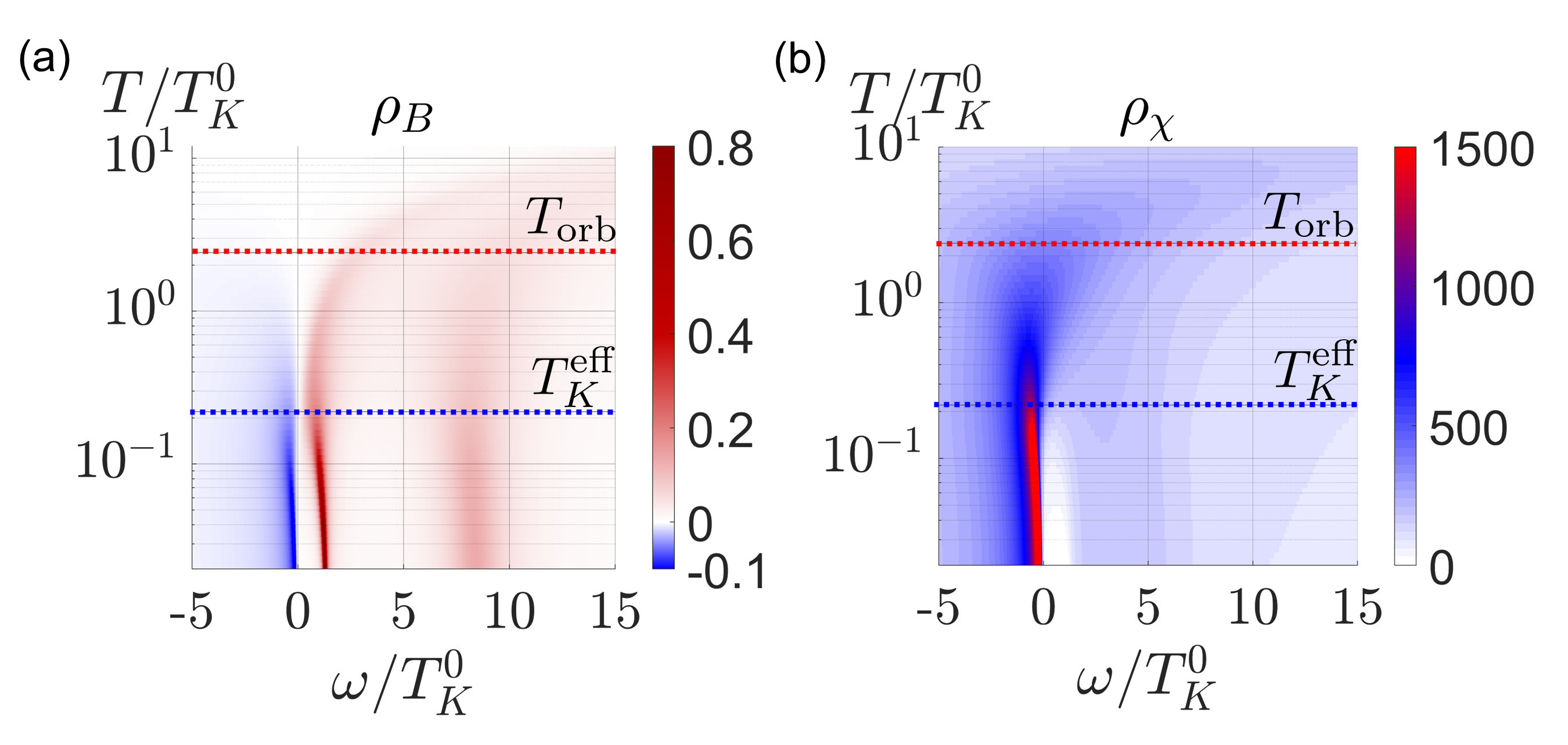}
    \caption{ (a) Temperature dependent  Schwinger boson's spectral function and (b) holon's spectral function for the intermediate regime.  In the locked moment phase, the low frequency peak moves downwards, finally settling into a constant value once $T$ reaches $T_K^{\rm eff} \sim 0.22 \: T_K^0$. This leads to a clear gap in the bosonic and holon spectra at low frequency, while the high frequency spinon band remains. Parameters used were $q = 0.3$ and $T_K^{0}/J_H = 0.08$, corresponding to a spinon gap of $\Delta/T_K^0 = 3.3$.}
    \label{fig:figure_33}
\end{figure}

In the intermediate regime, where $T_K^0/J_H < 1$ while $J_H \ll D$, we observe some of the most drastic effects of the Hund coupling on the holon and spinon's dynamics. {As temperature is reduced, the first effect that takes place is the settling of the Schwinger bosons into their lowest energy state. This is seen in the splitting in two of the high-frequency feature of the spinon's spectral function in two at $T_{\rm orb}$. These two features are separated by $3\Delta$, the upper band corresponding to excitations out of the fully ferromagnetically aligned state, and the lower one to energy fluctuations to the underscreened Kondo regime. The presence of a large spinon gap dampens the $k=\pm 2\pi/3$ spinonic excitation modes, and most of the spectral weigth moves to the $k=0$ mode as temperature is further reduced}. The low-frequency feature then keeps moving to lower frequencies, passing by the uncoupled system's gap edge $\Delta_K^{0}$. It is prevented from settling there since a significant amount of spectral weight still resides in the upper spinon band. Finally, as $T\simeq T_K^{\rm eff}$, a much smaller gap $\Delta_K^{\rm eff} \ll \Delta_K^{0}$ develops in both the spectral function $\rho_i = \frac{1}{\pi} \Imp G_i (\omega - i \eta)$ for both $G_{B, \rm loc}$ and $G_{\chi}$, as can be seen in Fig.~\ref{fig:figure_33}. The fixing and sharpening of the low-frequency mode is accompanied by another negative frequency resonance confining the spinons and indicating that the Kondo effect has fully settled in.

\begin{figure}[tb]
    \centering
    \includegraphics[width=\linewidth]{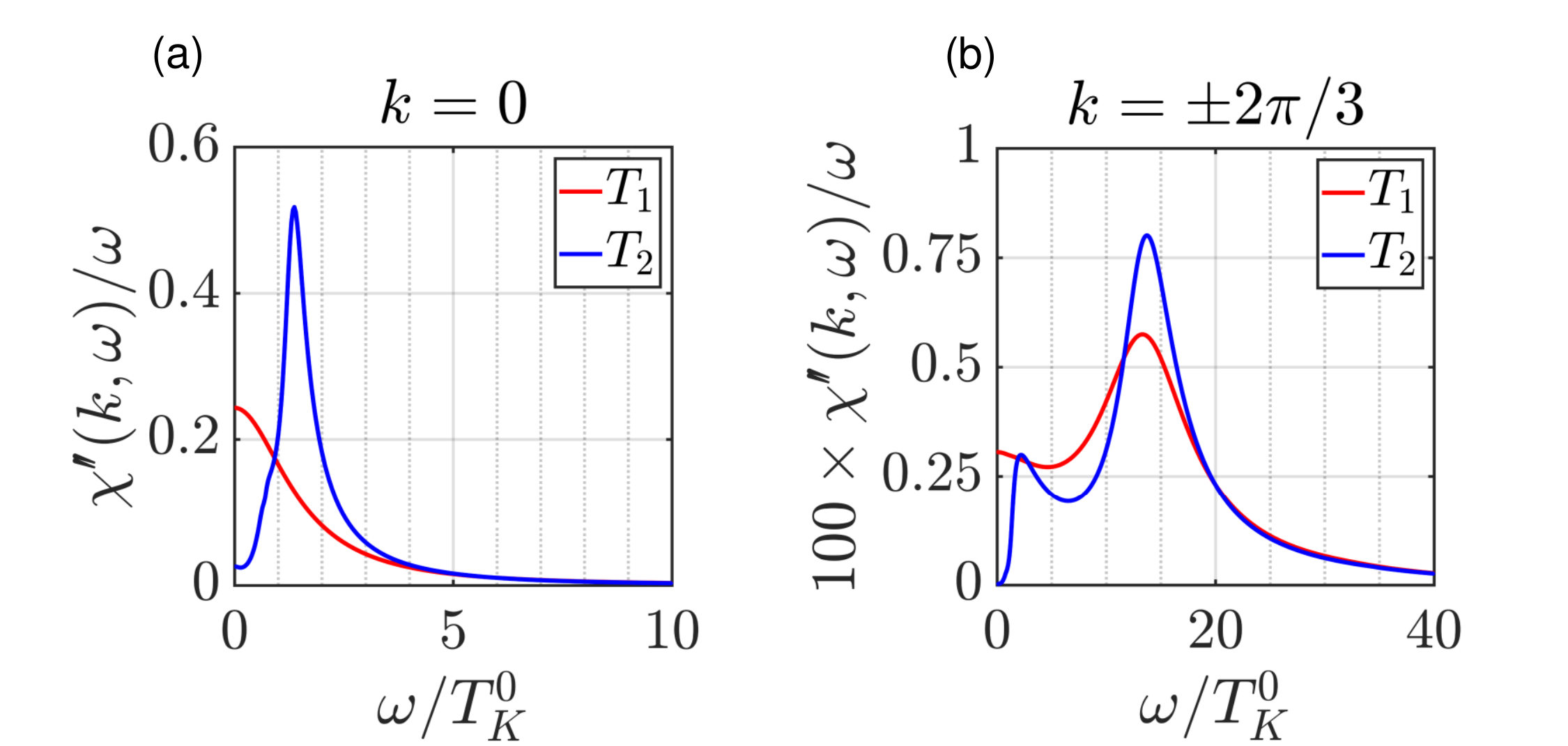}
    \caption{ Dynamical spin susceptibility $\chi'' (k, \omega)/\omega$ for temperature $T_1$ in the locked moment regime, and $T_2< T_K^{\rm eff}$ in the Fermi-Liquid regime, for $k=0$ (a) and $k=\pm 2\pi/3$ (b, enhanced by a factor of 100 for visibility). Parameters used were $q = 0.3$ and $T_K^{0}/J_H = 0.08$, corresponding to a spinon gap of $\Delta/T_K^0 = 3.3$.}
    \label{fig:figure_32}
\end{figure}

These effects are further evident when looking at the dynamical susceptibility, in Fig~\ref{fig:figure_32} (b). For a temperature $T_1$ below $T_{\rm orb}$ but much higher than $T_K^{\rm eff}$, the Kondo gap has not been established, and $\chi (0, \omega)$ presents a large maxima at zero frequency. However,  $\chi (2\pi/3, \omega)$ has a broad maxima at approximated $\omega \sim 3\Delta$, corresponding to spinon excitation out of the ferromagnetically ordered state. This peak is present for a large swath of temperatures, and remains prominent for $T_2 < T_K^{\rm eff}$. At these low temperatures, the low-energy gap starts to form and pushes the $k=0$ peak to finite frequency, opening a gap. Note that for the $T_2$ chosen, there remains a finite density of states in the gap, leading to a non-zero value of the $k=0$ dynamical susceptibility at zero frequency.  

For $J_H \gg T_K^0$, i.e. in the pure Schrieffer limit, solving the self-consistency equations becomes more and more unstable at high-temperature. In this limit, the high-frequency spinon excitation band moves completely out of the electronic bandwidth, such that effectively, $G_{B, \rm loc}(z) \sim \frac{1}{3} G_B (p=0, z)$. This amounts a condensation of the spinons into the uniformly aligned state at high-temperature, akin the RG flow from the locked moment fixed point of Fig.~\ref{fig:figure_1}. Since the moment is already formed, further increasing $J_H$ does not change $T_K^{\rm eff}$. We find that $T_K^{\rm eff}/D \sim (T_K^0/D)^{2.6}$ in the Schrieffer limit, as opposed to the expected $(T_K^0/D)^3$, using the maximum of the spin susceptibility as our indicator for $T_K^{\rm eff}$. An alternate way to indicate the Kondo temperature is through the progression of the holon phase shift towards unitarity: $\delta_{\chi} \rightarrow \pi$. If we set $\delta_{\chi}/\pi = a$ at $T = T_K^{\rm eff}$, then the obtained exponent for the Schrieffer limit tends towards the expected $3$ as $a$ is taken closer and closer to $1$, i.e. full unitary phase shift. We view this as a confirmation that we retrieve the Schrieffer limit with the full numerical treatment of the  large-N equations.

It is clear that one could change $M$ and repeat this exhaustive investigation of the exponential depletion of $T_K^{\rm eff}$. We note that already for $M=3$, the effective Kondo temperature for large $J_H$ is already more than 4 orders of magnitude decreased. For realistic systems with $M=5$, for example Mn${}^{2+}$, the effective Kondo temperature would be more than 10 orders of magnitude below. The trend from Fig.~\ref{fig:figure_2} would be downward with an ever increasing slope for larger $M$. Even at $M=5$, the physics would look very similar to that of a ferromagnetic 1D chain \cite{komijani2018model}, with the effective Kondo temperature so low that it would be effectively un-measurable experimentally.


\section{Kondo Temperature and single-iteration approach} \label{sec4}

An approximate treatment of the effective Kondo temperature can be obtained with only a single iteration of the self-consistency equations. 

\subsection{Kondo temperature}

We first establish a simple analytical criterion for the Kondo temperature. 

As the temperature is lowered from high temperatures, a pole in the holon's Green's function develops and moves from positive frequency to negative frequency, 
where a bound state forms. This consideration provides an estimate for the Kondo temperature as the temperature at which the holon pole is at zero frequency. 
Because of the structure of the holon Green's function from equation~\eqref{eq:holonG}, this leads to 

\begin{equation}
    -\frac{1}{J_K} = \Rep \Sigma_{\chi} (\omega = 0+ i \eta) \qquad \text{at} \qquad T = T_K. \label{eqOneLoop}
\end{equation}

Translating Eq.~\eqref{eq:selfcons} from imaginary time $\tau$ to real frequency $\omega$, and then taking $\omega \rightarrow 0$, we get the following equation which implicitly defines $T_K$

\begin{align}
    -\frac{1}{J_K} = \int \frac{d \omega}{\pi} & \left[G_{B, \rm loc}'' (\omega) n_B(\omega) g_c'(\omega) \right. \notag \\
    &  \left. -G_{B, \rm loc}'(\omega) f(\omega) g_c''(\omega) \right] .
\label{eq:conditionSI}
\end{align}

\subsection{Single-iteration approach}

We evaluate this definition of the Kondo temperature in a single-iteration approximation, which we introduce here. We first solve the saddle point equations~\eqref{eq:qDelta} using the bare local bosonic Green's function, i.e. equation~\eqref{eq:bosonG} with $\Sigma_B = 0$. After slightly changing the notation by using $\lambda' = \lambda - 2\Delta$, we obtain

\begin{equation}
\begin{split}
3\Delta &= J_H [n_B (\lambda') - n_B(\lambda' + 3\Delta)] \\
3 q &= [n_B (\lambda') + 2 n_B(\lambda' + 3\Delta)].
\end{split} \label{eqs:qDelta2}
\end{equation}

\begin{figure}[h]
    \begin{center}
    \includegraphics[width=0.9\columnwidth]{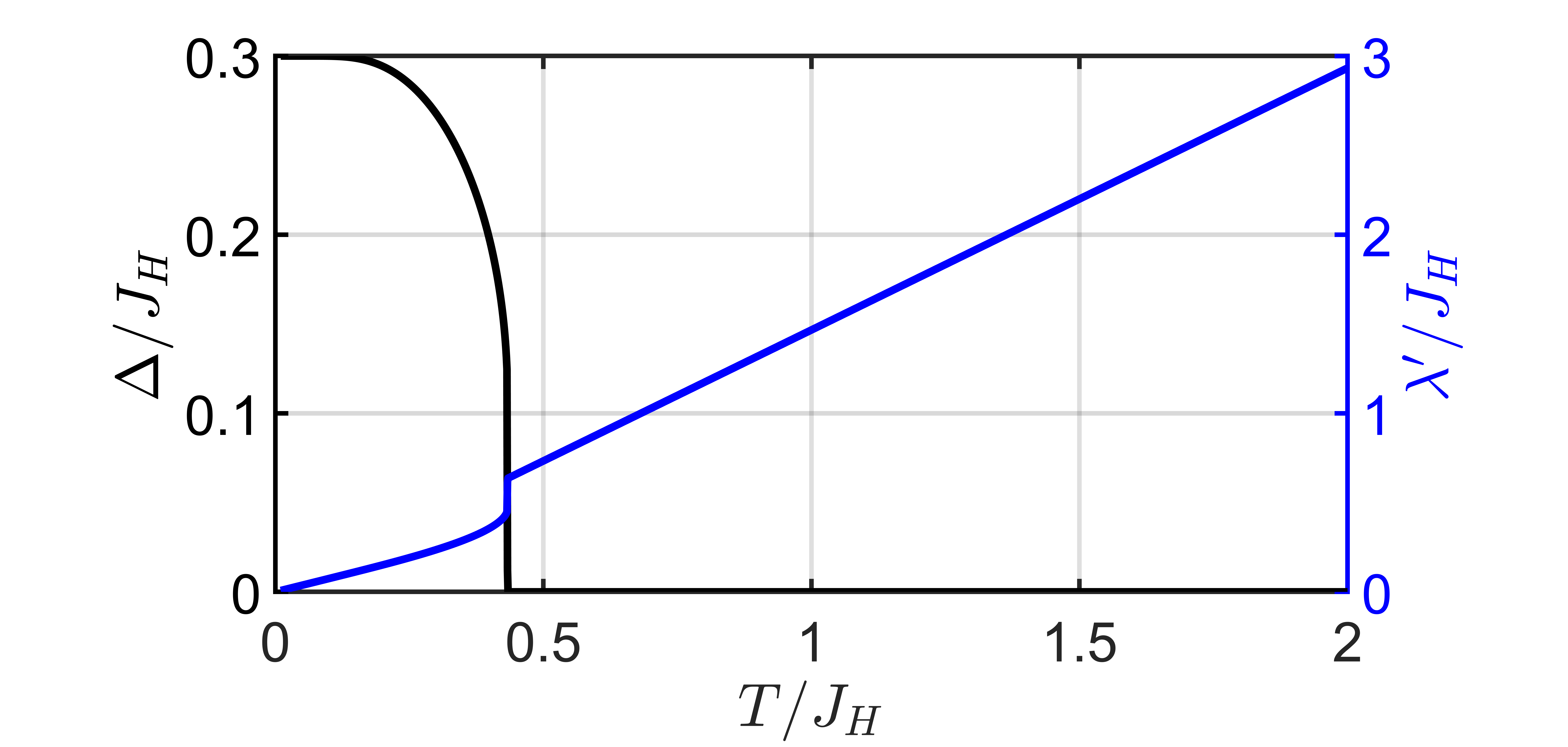}
    \end{center}
    \caption{Spinon gap $\Delta$ (left) and modified spinon chemical potential $\lambda' = \lambda - 2 \Delta$ (right) obtained from equations~\eqref{eqs:qDelta2}, as a function of temperature. Parameters used were $q = 0.3$.}
    \label{fig:DeltaLam}
\end{figure}

Solving this system of equations, we find $\lambda'$ and $\Delta$ versus temperature $T$ for different values of $s$ and $J_H$, Fig.~\ref{fig:DeltaLam}. For temperatures above a certain $T_{\rm orb}$, the spinon gap $\Delta$ is $0$. There, the spinon chemical potential is the same as it would be for a single impurity, i.e. $\lambda_{\text{high T}} = T \log (1 + 1/q)$. At low temperature on the other hand, the gap fully develops, and one gets that $\Delta \sim q J_H$ and $\lambda' \sim T \log (1 + \frac{1}{3q})$. This is consistent with the formation of the large moment of size $q' = 3q$. 

\vspace{3mm}

\subsection{Approximate Kondo temperature} \label{subsecTKsi}

Here, we estimate $T_K$ as defined through Eq.~\eqref{eq:conditionSI} using the approximate single-iteration solution, Fig.~\ref{fig:DeltaLam}. The first line of Eq.~\eqref{eq:conditionSI}, $\int d\omega G_{B, \rm loc}'' (\omega) n_B(\omega) g_c'(\omega)/\pi$, describes the correction to the holon self-energy due to an on-shell spinon and virtual conduction electron. In contrast, the second line, $\int d\omega G_{B, \rm loc}'(\omega) f(\omega) g_c''(\omega)/\pi$, describes the reverse process: here the conduction electron is on-shell, while the bosonic spinon is virtual. Clearly, the phase space for the second process is parametrically larger, because the conduction electrons form a continuum, while the spectrum of bosons in the limit of $\Sigma_B = 0$ is discrete. As a consequence, see Appendix~\ref{appendix2}, in the limit of $D/T \rightarrow \infty$, the integral in the first line of Eq.~\eqref{eq:conditionSI} is negligible compared to the second line, which itself results in Eq.~\ref{eq:SIv1}~\cite{coleman2015introduction}. In this equation, $P.V.$ is the principal value, and $\tpsi (z) = \Rep \psi(1/2- i \:z/2\pi)$, where $\psi(z)$ is the digamma function. We here considered only the realistic case of $\lambda' \ll \min \{D, J_H\}$. In Eq.~\eqref{eq:SIv1}, we introduced a soft cutoff on the conduction electrons $D^2/(D^2 + \omega^2)$ instead of the sharp cutoff at $\omega = \pm D$.

\begin{widetext}
\begin{align}
-\frac{3}{J_K \rho} &= P.V. \int_{-\infty}^{\infty} d\omega \frac{D^2}{\omega^2 + D^2} \left(\frac{2 f(\omega)}{\omega - 3\Delta - \lambda'} + \frac{f(\omega)}{\omega  - \lambda'}\right) \notag \\
&=    \ln(2\pi T/D)  + \tpsi(\lambda'/T)  + \frac{2}{1 + 9\Delta^2/D^2} [\:  \ln(2\pi T/D) + \tpsi(\lambda'/T + 3\Delta/T) \:] - \frac{\pi}{D/3\Delta + 3 \Delta/D}. \label{eq:SIv1}
\end{align}
\vspace{5mm}
\end{widetext}

\begin{figure}[h]
    \centering
    \includegraphics[width=\linewidth, trim=4 4 4 4,clip]{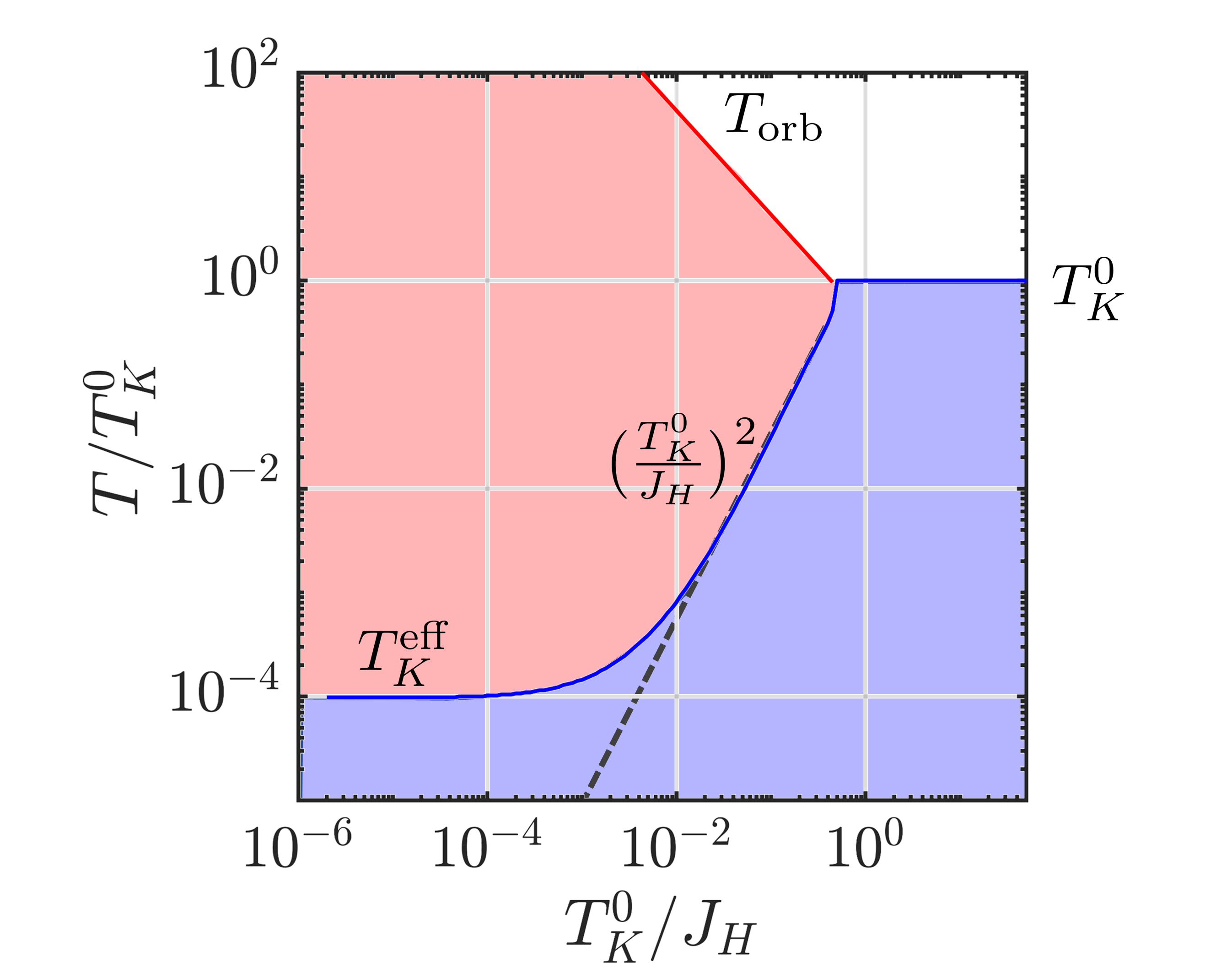}
    \caption{ Phase diagram obtained from a single iteration of the large-N equations, by solving Eq.~\eqref{eq:SIv1}, for $q = 0.3$, $D = 300$ and $T_K^0/D = 0.007$. The blue line is the solution $T_K^{\rm eff}$, while the red line $T_{\rm orb}$ is obtained from the solution of Eq.~\eqref{eqs:qDelta2} for a given $s$ and $J_H$. The black dashed line corresponds to the scaling at intermediate $T_K^0/J_H$.}
        \label{fig:PhaseDiagram2}
\end{figure}

This is a nonlinear equation which is then solved assuming the $\lambda'(T)$ and $\Delta(T)$ behavior from the solution of Eq.~\eqref{eqs:qDelta2} for a given $J_K$, $\rho$, $J_H$ and $q$. The temperature that satisfies this equation is then $T_K^{\rm eff}$ as obtained by the single iteration method.  This semi-analytical formalism reveals outstanding details in some limiting cases.  The result is shown in Fig.~\ref{fig:PhaseDiagram2}.

The single impurity result $T_K^0$ is easily obtained by setting $\Delta = 0$ and $\lambda' = T \lambda_0$ with $\lambda_0 \sim \log (1 + {{1}/{q}})$. This leads to  

\begin{equation}
\begin{split}
\frac{T_K^0}{D} &\simeq \frac{1}{2\pi} \exp{\left[ -\frac{1}{J_K \rho} - \tpsi( \lambda_0)\right]}, 
\end{split}
\end{equation}

\noindent retrieving the well-known result. We use this expression to obtain a compact self-consistent condition for $T_{K}^{\rm eff}$ in the limit $D \gg J_H, T_K^0$,

\begin{equation}
\frac{T_K^{eff}}{D} = \frac{T_K^{0}}{D} \exp{[\tpsi( \lambda_0) -  \frac{1}{3}\tpsi( \tilde{\lambda}) - \frac{2}{3}\tpsi( \tilde{\lambda} + 3 \tilde{\Delta}) ] } \; , 
\end{equation}

\noindent with $\tilde{\lambda} = \lambda'(T_K^{\rm eff})/T_K^{\rm eff}$ and $\tilde{\Delta} = \Delta(T_K^{\rm eff})/T_K^{\rm eff}$. 

For $J_H \gg T_K^0$, we can simplify the result much further using $\Delta \simeq q J_H$, $\lambda' \simeq T \log(1 + {{1}/{3q}})$ and $\psi(z) \sim \log(z)$ for large $z$ arguments

\begin{align}
\frac{T_K^{\rm eff}}{D} & \sim  \frac{T_K^{0}}{D} \left( \frac{T_K^{\rm eff}}{3qJ_H} \right)^{2/3} ,
\end{align}

or, equivalently, for $D \gg J_H \gg T_K^0$

\begin{equation}
\frac{T_K^{\rm eff}}{T_K^{0}} \sim \left( \frac{T_K^{0}}{3qJ_H} \right)^{2},
\end{equation}

\noindent which is identical to the form of Ref.~\cite{nevidomskyy2009kondo} that was obtained using a renormalization group study of the SU(2) Hund-Kondo model.

A final important limit to mention is one with $J_H \gg D \gg T$. In that case, the spinon band is excluded from the electron bandwidth, and the system behaves right away like one of $q' = 3q$. By examining equation~\eqref{eq:SIv1}, it is clear that while $\lambda'$ remains proportional to $T_K^{\rm eff}$, $\Delta$ is very large due to the large $J_H$. Hence, the second term of \eqref{eq:SIv1} tends to $0$ in that limit where $D/\Delta \rightarrow 0$. The resulting self-consistent equation is then

\begin{equation}
\frac{T_K^{\rm eff}}{D} = \frac{1}{2\pi} \exp{\left[ -\frac{3}{J_K \rho} - \tpsi( \lambda_S) \right]} \propto  \left( \frac{T_K^{0}}{D}\right)^3 \;,
\end{equation}

\noindent with $\lambda_S = \log(1 + \frac{1}{3q})$ so that one obtains the identical Schrieffer limit for very large Hund's coupling. The resulting full $T_K^{\rm eff}$ curve obtained by solving equation~\eqref{eq:SIv1} is presented in figure~\ref{fig:PhaseDiagram2}. One can see that even at the single-iteration level, the exponential depletion of $T_K^{\rm eff}$ in the presence of $J_H$ is correctly captured. Furthermore, both the exponent $\beta = 2$ of that decrease in the regime of intermediate $J_H$  and the Schrieffer limit for large $J_H \gg D$ are appropriately conveyed with this approach.  


Finally, we note the striking qualitative similarity with the phase diagram obtained from the full numerical solution of the large-N self-consistent equation, presented in Fig.~\ref{fig:figure_2}. The essential difference between them is twofold. Firstly, we used here a very soft cutoff for the conduction electron, compared to a sharper one in our numerical phase diagram. Secondly, self-energy effects for the spinons change the value of $T_K^0/J_H$ where the downward renormalization starts to take effect.


\section{Conclusions and outlook} \label{secEnd}

We here present the main conclusions of this work, as well as future directions for the large-N method in multiorbital systems.

\subsection{Summary}

The strength of the large-N Schwinger boson approach lies in both its simplicity and its ability to capture the essential features of a physical model, including its thermodynamic, dynamic and ground state properties. The analytical control of the method comes at the price of a large-N limit, so that the reader might wonder about the ability to correctly describe the physical SU(2) case. Importantly, we find that all the relevant regimes are both qualitatively and quantitatively retrieved. This is in line with previous studies of this method on both ferromagnetic and antiferromagnetic spin chains~\cite{komijani2018model, komijani2019emergent}. In particular, and in contrast to large-N Abrikosov fermion methods, the present theory correctly captures the emergence of Kondo screening as a crossover, rather than a second order phase transition.

The essence of the renormalization of the Kondo temperature in the presence of Hund's coupling is rather simple: as the local moments bind ferromagnetically, they form a larger moment which exponentially suppresses the formation of the Kondo-singlet. Through both our numerical solution of the self-consistent self-energy equations and our single-iteration approach, we were able to correctly capture the exponential decrease of $T_K^{\rm eff} \sim (T_K^0/J_H)^2$, in alignment to previous RG work~\cite{nevidomskyy2009kondo}, as well as the Schrieffer limit where $J_H > D$. 

One of the successes of this approach is the ability to compute thermodynamic observables with accuracy. In particular, the impurity entropy and specific heat are extracted using an exact formula in the large-N limit. Comparing the thermodynamical quantities to analytical limits in the free moments and locked moments regimes showed good agreement, as well as the clear evidence of the emergence of the large moments. While this is associated with a loss of entropy at the impurity site, our calculations of the dynamical susceptibility reveal a clear feature at high-frequency corresponding to spinon fluctuations. Such excitations remain extremely short lived, protecting much longer timescales than the bare inverse Kondo time $1/T_K^0$. This connects with the idea of long-lived moments in the Hund's coupled impurity. 

\subsection{Outlook} \label{sec:outlook}

We conclude with two perspectives for future research. First, in regards of Hund's impurity models, a recent body of work using numerical renormalization group (NRG) has demonstrated the proximity of the doped multi-orbital Kondo model to a non-Fermi-liquid (nFL) fixed point \cite{horvat2019non,walter2020uncovering, wang2020global}. In the case where two electrons occupy three orbitals, the model bears a striking resemblance to the overscreened Kondo model, which has a nFL fixed point. The influence of the latter has been revealed upon tuning the interactions between the orbitals. The natural extension of the work presented here more relevant to the Fe$^{2+}$ configuration is to consider mixed valence scenarios, in which the holons are real particles rather than virtual particles mediating Kondo interaction. Such an infinite-U Hund-Anderson impurity model is expected to display richer interplay between Hund's interaction and the Kondo physics.

\begin{figure}
    \centering
    \includegraphics[width=\linewidth]{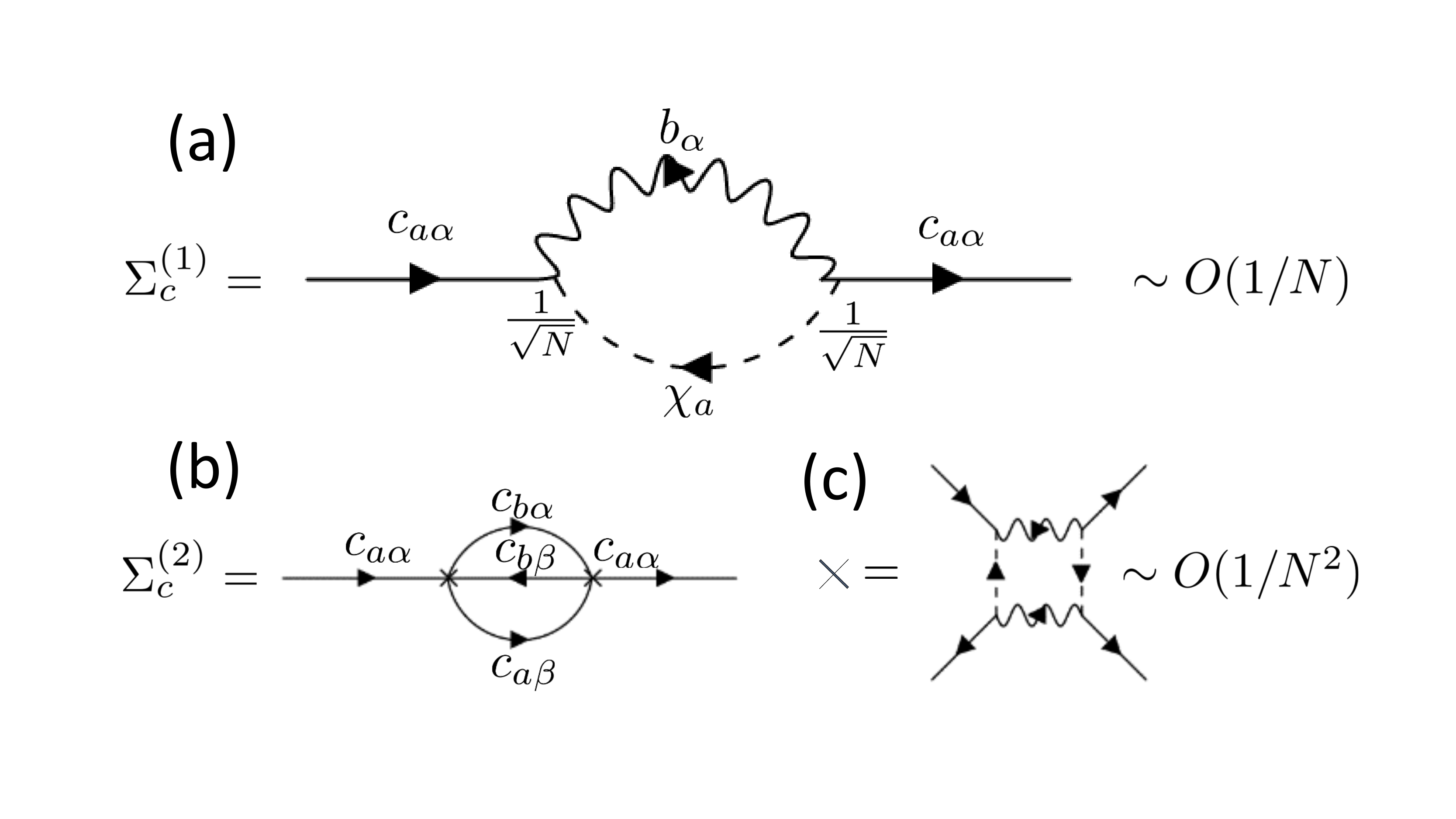}
    \caption{
    (a) The electronic self-energy $\Sigma_{c}^{(1)} = \frac{1}{N} G_{\chi}(-\tau) G_B (\tau)$ (we suppressed the orbital index $m$) obtained from the Luttinger-Ward functional. (b) Next-order correction which includes inelastic scattering of the conduction electrons off the impurity, where vertices are represented in (c).}
    \label{fig:figure_5}
\end{figure}

In closing, we remark that the exponential suppression of the Fermi liquid temperature due to Hund's coupling, re-established here, may be linked to the apparent direct transition from non-Fermi liquid to superconducting state seen in iron-based superconductors.

Beyond the problem of an isolated impurity, a second perspective is to use the present Schwinger boson formalism as a real-frequency impurity solver for a dynamical mean-field theory (DMFT) of extended Hund's metals. Technically, this question is intimately linked to the self-energy of conduction electrons, which in the present approach scales as $1/N$ in the large-N limit, as opposed to the spinon and holon self-energies which are $O(1)$ (see Fig.~\ref{fig:figure_5} (a)). In a DMFT spirit, one may employ $\Sigma_c$ self-consistently with the other self-energies, keeping $N$ finite. This was done in references \cite{lebanon2006conserving, lebanon2007fermi} for an impurity with a single orbital, where a (local) Fermi-liquid with a finite electronic phase shift was observed. It is worth asking however if the signature of the Fermi liquid, the $T^2$ resistivity, can be recovered using this method. Unfortunately, using $\tilde{\Sigma}_c = N\Sigma_c$ to leading order in $N$, the expected $T^2$ behavior is replaced by an exponential decay due to finite gap in both the holons and spinons \cite{lebanon2007fermi, wang2020quantum}. We envision that one could use the obtained self-consistent Green's functions into $\Sigma_c^{(2)}$ (see Fig.~\ref{fig:figure_5} (b)) to restore the inelastic scattering of the electrons off the impurity. Taken together, these next steps together would make the large-N approach a viable alternative to multi-orbital impurity solvers by providing an accurate electronic self-energy.

\section*{Acknowledgments} \label{secAcknowledge}

This work was supported by DOE Basic Energy Sciences grant DE-FG02-99ER45790 (VDT, EJK, PC ) and NSF grant DMR-1830707 (YK)
and the Fonds de Recherche Québécois en Nature et Technologie (VDT).

\section*{Appendices}

\appendix

\section{Mean field solution at $J_K = 0$} \label{appendix1}

\begin{figure}[h]
    \centering
    \includegraphics[width=0.6\linewidth]{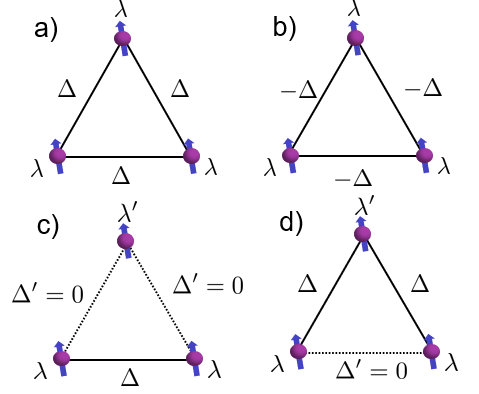}
    \caption{The four examined mean-field scenarios for the Hund's triangle. In order, these are the homogeneous solution with positive spinon hoping (a), with negative spinon hopping (b), the dimerized solution where only 2 sites are active with each other forming a dimer in (c), and the situation where one link is missing.}
    \label{fig:app_1}
\end{figure}

We consider the Hund-Kondo Hamiltonian using the Schwinger boson representation of the orbital moments, in the absence of Kondo coupling. The most general mean field model of the Hund coupling in this situation is

\begin{align}
H_{\rm MFT} &=  \begin{pmatrix}
b^{\dagger}_{1} & b^{\dagger}_{2} & b^{\dagger}_{3}
\end{pmatrix}
\begin{pmatrix}
\lambda_1 & \Delta_1 & \Delta_3\\
\Delta_1 & \lambda_2& \Delta_2\\
\Delta_2 & \Delta_3 & \lambda_3
\end{pmatrix}
\begin{pmatrix}
b_{1} \\ b_{2} \\ b_{3}
\end{pmatrix} \notag \\
&+ \sum_i \left( \frac{|\Delta_i|^2}{J_H} - \lambda_i q \right) \; ,
\end{align}

\noindent with $q = 2S/N$, and where we consider the most general case where the $\lambda_i$ can all be different, and $\Delta_i =: \Delta_{i, i+1}$ with periodic boundary condition. We diagonalize the matrix above for different scenarios of $\lambda_i$ and $\Delta_i$, and evaluate the free energy $f = \frac{F}{N} = T\sum_i \log (1 - e^{-E_i/T}) + \sum_i \left( \frac{|\Delta_i|^2}{J_H} - \lambda_i q \right)$ under the stationarity conditions $\partial_{\lambda} f = 0$ and $\partial_{\Delta} f = 0$. The four scenarios are shown in figure~\ref{fig:app_1}. 

Diagonalization of the mean-field hamiltonian reveals the following spectra for the 4 configurations examined:

\begin{equation}
\begin{split}
E_a &=  (\lambda - \Delta, \lambda - \Delta, \lambda + 2\Delta) \; , \\
E_b &= (\lambda - 2 \Delta, \lambda + \Delta, \lambda + \Delta) \; , \\
E_c &=  (\lambda - \Delta, \lambda + \Delta, \lambda') \; , \\
E_d &= (\frac{1}{2} ((\lambda + \lambda') - \sqrt{(\lambda - \lambda')^2 + 8 \Delta^2} ), \\
& \; \lambda', \frac{1}{2} ((\lambda + \lambda') + \sqrt{(\lambda - \lambda')^2 + 8 \Delta^2} ) ) \; .
\end{split}
\end{equation}

For configuration (a), the stationarity conditions with respect to $\lambda$ and $\Delta$ lead to 

\begin{equation}
\begin{split}
3q &=  2n_B(\lambda - \Delta) + n_B(\lambda + 2 \Delta) \; , \\
\frac{3\Delta}{J_H} &= n_B(\lambda - \Delta) - n_B(\lambda + 2\Delta) \; ,
\end{split}
\label{eq:app1c1}
\end{equation}

\vspace{2mm}

\noindent where $n_B(x) = 1/( e^{x/T} - 1)$. At low temperature, assuming $\lambda-  \Delta >0$ (keeping this the lowest energy state), then $n_B(\lambda-\Delta) \simeq T/(\lambda- \Delta) $ and $n_B(\lambda + 2\Delta) \simeq 0$. From Eq.~\ref{eq:app1c1}, we find that this leads to $\Delta = q J_H$ and then that $\lambda - \Delta = \frac{2T}{3q}$. Putting this back into the free energy, and approximating that $\ln (1 - x) \simeq -x$ at low temperatures, then $f_a \rightarrow 0$.

For configuration (b), the stationarity conditions with respect to $\lambda$ and $\Delta$ lead to 

\begin{equation}
\begin{split}
3q &=  2n_B(\lambda + \Delta) + n_B(\lambda - 2 \Delta)\; , \\
\frac{3\Delta}{J_H} &= n_B(\lambda - 2 \Delta) - n_B(\lambda + \Delta) \; .
\end{split}
\label{eq:app1c2}
\end{equation}

\noindent Keeping $\lambda - 2\Delta$ as the lowest energy state as $T \rightarrow 0$, this leads to very similar conditions to (a) where now $\lambda - 2\Delta = \frac{T}{3q}$ and $\Delta = q J_H$, we get $f_a \rightarrow  -3 J_H q^2$ .

For configuration (c), the dimerized setup, there are now three stationarity conditions, as $\lambda$ and $\lambda'$ are now tunable parameters. These lead to 

\begin{equation}
\begin{split}
2q &=  n_B(\lambda - \Delta) + n_B(\lambda +  \Delta) \; ,\\
\frac{2\Delta}{J_H} &= n_B(\lambda - \Delta) - n_B(\lambda + \Delta)\; , \\
q &= n_B(\lambda')\; .
\end{split}
\label{eq:app1c3}
\end{equation}

\noindent Keeping $\lambda - \Delta$ as the lowest energy state as $T \rightarrow 0$, this leads $f_c \rightarrow  - q^2 J_H$.

The same procedure for configuration (d) leads to a cumbersome set of equation. In the low-temperature limit, these lead to $\Delta = \sqrt{2} J_H q$ and $\lambda - \sqrt{2}\Delta = \frac{T}{4q}$, which gets a free energy of $f_d \rightarrow (4 - 3\sqrt{2}) J_H q^2$. 

The calculation of the low-temperature free energy shows that at low-temperatures, configuration (b) is the lowest energy state, and we are correct in assuming $\Delta_m = -\Delta$ and $\lambda_m = \lambda$ for the model with $J_K$ finite. This concludes the derivation of the Mean-Field solution used in bosonic Green's function at equation~\eqref{eq:bosonG}.

\section{Entropy and susceptibility - limiting cases} \label{appendix3}

At finite temperature, our free-energy $f(T) = F(T)/N$  only has a few terms that remain $\text{O}(1)$. We obtain the following form for $k = K/N = q = 2S/N$

\begin{widetext}
\begin{align}
    f(T) &=  \sum_p \int \frac{d\omega}{\pi} n_B (\omega) \left[ \Imp \log (- G_{B}^{-1} (p, \omega)) + \Sigma_B'' (\omega) G_{B}' (p, \omega) \right] \nonumber\\
    & \; \; - q \int \frac{d\omega}{\pi} n_F (\omega) \left[ \Imp \log (- G_{\chi}^{-1} (\omega)) + \Sigma_{\chi}'' (\omega) G_{\chi}' (\omega) + \tilde{\Sigma}'_c (\omega) g_{c,0}'' (\omega) \right]   - 3\lambda q   \label{eq:app3v1} \; , \\
    S_{\rm imp}(T) &= - \sum_p \int \frac{d\omega}{\pi} \partial_T n_B (\omega) \left[ \Imp \log (- G_{B}^{-1} (p, \omega)) + \Sigma_B'' (\omega) G_{B}' (p, \omega) \right] \nonumber\\
    & \; \; - q \int \frac{d\omega}{\pi} \partial_T n_F (\omega) \left[ \Imp \log (- G_{\chi}^{-1} (\omega)) + \Sigma_{\chi}'' (\omega) G_{\chi}' (\omega) - \tilde{\Sigma}'_c (\omega) g_{c,0}'' (\omega) \right] \label{eq:app3v2} \; ,
\end{align}
\end{widetext}

\noindent where $\tilde{\Sigma}_c = N \Sigma_c = G_{\chi}(-\tau) G_B (\tau)$. The total impurity entropy is obtained from this using $ S_{\rm imp} = - \frac{\partial f}{\partial T}$. One can then evaluate this in different regimes. 

At high-temperatures ($T \gg \Delta$), we can neglect all self-energy contributions. This leads to $G_B (p = 0, \omega) \simeq [\omega - \lambda + 2\Delta]^{-1}$ and $G_B (p = \pm 2\pi/3, \omega) \simeq [\omega - \lambda - \Delta]^{-1}$, such that the spectral functions are $\rho_B(q,\omega) = \delta(\omega - \lambda - \epsilon_q)$. Since the constraint is always enforced, we get that $q = \int d\omega n_B(\omega) \rho_B(\omega) = \frac{1}{3}n_B (\lambda - 2\Delta) + \frac{2}{3}n_B (\lambda + \Delta)$. In the limit $T \gg \Delta$, we have that $n_B (\lambda - 2\Delta) \simeq n_B (\lambda + \Delta) \simeq n_B(\lambda)$. This leads to ${q}= n_B(\lambda^{\text{High T}})$, which, if inverted, give the high-temperature value of the spinon's chemical potential.

\begin{equation}
    \lambda^{\text{High T}} = T\log \left( \frac{1 + q}{q} \right) \; ,
\end{equation}

\noindent We insert this into the impurity's total free energy.

\begin{equation}
\begin{split}
    f^{\text{High T}}_{\text{imp}} &= T \sum_p \sum_n \log (\beta(-i\nu_n + \lambda^{\text{High T}} + \epsilon_p))   \notag \\ & - 3q \lambda^{\text{High T}} (T) \\
    &= T \ln(1 - e^{-\beta(\lambda^{\text{High T}} - 2\Delta)}) \notag \\ & +  2T \ln(1 - e^{-\beta(\lambda^{\text{High T}} + \Delta)})  \notag \\ & -3T[q \log (1 + q) - q \log (q)] \\
    &= -3T[(1 + q)\log (1 + q) - q \log (q)] \; ,
\end{split}
\end{equation}

\noindent so that the high-temperature entropy limit is extracted from $ S_{\rm imp} = -\partial_T f(T) = 3[(1 + q)\log (1 + q) - q \log q]$. One can also take an alternative limit where $\Delta \gg T$ at still high-temperature. The spectral function is the same, but this leads to a new consequence of the Schwinger boson number constraint. In this limit, we have that $n_B(\lambda + \Delta) \rightarrow 0$, and $n_B (\lambda - 2\Delta)$ remains finite. This leads to ${q}= \frac{1}{3}n_B(\lambda^{\text{High } \Delta} - 2\Delta)$. Solving this leads to 

\begin{equation}
    \lambda^{\text{High }\Delta} = 2\Delta + T\log \left( \frac{1 + 3q}{3q} \right) \; ,
\end{equation}

In effect what we have here is an effective $q$ such that $q' = 3q$. Calculating the free energy, we get 

\begin{widetext}
\begin{equation}
\begin{split}
    f^{\text{High }\Delta}_{\text{imp}} &= T \sum_p \sum_n \log (\beta(-i\nu_n + \lambda^{\text{High }\Delta} + \epsilon_p)) - 3q  \lambda^{\text{High }\Delta} (T) \\
    &= T \ln(1 - e^{-\beta(\lambda^{\text{High }\Delta} - 2\Delta)})  -3qT[ \log (1 + 3q) -  \log (3q)] - 6q \Delta \\
    &= -T[(1 + 3q)\log (1 + 3q) - 3q \log (3q)] - 6 q \Delta \; ,
\end{split}
\end{equation}
\end{widetext}

\noindent and hence the impurity entropy is $ S_{\rm imp} = -\partial_T f(T) = [(1 + 3q)\log (1 + 3q) - 3q \log 3q]$. This is a rather large decrease in entropy between the truly high-temperature regime and the high-$\Delta$ regime (corresponding to the onset of Hund's coupling and ferromagnetic order in the triad). We expect that in situations where the three regimes are well separated, these two limits will be present as entropy plateaus, and the transition between them will be at $T_{\rm orb}$ corresponding to a large specific heat peak. 

We now focus on the same limiting procedure for the susceptibilities. An important form we will need for these derivation is $\Imp G(\omega)^2$. Let $G (\omega) = [\omega - \Omega]^{-1}$. Then, $\Imp G^2(\omega) = 2 \pi \delta'(\omega - \Omega)$ where $\delta'(x) = \partial_x \delta(x)$. This statement is important in the following derivations.

We recall the overall forms for the susceptibilities:

\begin{align}
    \chi_0 &= \int \frac{d\omega}{\pi} n_B (\omega) \Imp [\chi_{0,\omega}] \; , \nonumber \\
    \chi_{0,\omega} &= \frac{1}{3} \sum_p G_B^2 (p, \omega )\; ,  \\
    \chi_{\pm} &= \int \frac{d\omega}{\pi} n_B (\omega) \Imp [\chi_{\pm,\omega}] \; , \\
    \chi_{\pm,\omega} &= \frac{2}{3} G_B (0, \omega ) G_B (2\pi/3, \omega) + \frac{1}{3}G_B^2 (2\pi/3, \omega)  \; , \nonumber \\
    \chi_{loc} &= \int \frac{d\omega}{\pi} n_B(\omega) \Imp G_{B,\rm loc}^2 (\omega) = \frac{1}{3}[\chi_0 + 2 \chi_{\pm}]\; ,
\end{align}

\noindent with the local bosonic Green's function $G_{B, \rm loc} (\omega) = \frac{1}{3} \sum_p G_B (p, \omega)$. Starting then with $\chi_0 $ and neglecting all self-energy contributions, we get the expression in Eq.~\ref{eq:chi0app}. In the $T \gg \Delta$ limit, we can use the same asymptotic limit for $\lambda$, as well as the approximation that $n_B(\lambda + \Delta) = n_B (\lambda - 2 \Delta) = n_B(\lambda) = q$. This leads to $\chi_0 \rightarrow \frac{2q (1 + q)}{T}$ be Curie-like at high-temperature, with a moment size determined by $q = 2S/N$, as expected. In the $\Delta \gg T$ limit, the limiting $\lambda$ changes form to accommodate the effective $q' = 3 q$, and we have $n_B(\lambda + \Delta)  = 0$ and  $n_B (\lambda - 2 \Delta) = 3q$. This leads to $\chi_0 \rightarrow \frac{2q (1 + 3q)}{T}$ in that limit, in line with the idea of an emergent moment three times the size of the original local moment on the orbitals. 

\begin{equation}
\begin{split}
\chi_0 &= \frac{2}{3} \sum_p \int d\omega n_B (\omega) \delta'(\omega -  \lambda + 2\Delta \cos{p}) \\
&= -\frac{2}{3} \sum_p \int d\omega n_B' (\omega) \delta(\omega -  \lambda + 2\Delta \cos{p}) \\
&= \frac{2}{T}\left[\frac{2}{3} n_B (\lambda + \Delta) (1 + n_B (\lambda + \Delta)) \right. \\
& \; \; \left. + \frac{1}{3} n_B (\lambda - 2\Delta) (1 + n_B (\lambda - 2\Delta)) \right] \; .
\end{split}
\label{eq:chi0app}
\end{equation}

The same procedure can be done for $\chi_{\pm}$. Going back to our definition of the finite momentum susceptibility, we have that $\chi_{\pm} = \int \frac{d\omega}{\pi} n_B (\omega) \Imp [\chi_{\pm,\omega}]$ with $\chi_{\pm,\omega} = \frac{2}{3} G_{0} G_{\pm} + \frac{1}{3}G_{\pm}^2$ where $G_{q} = G_B (q, \omega + i \eta)$ and $\pm$ refers to $p = \pm 2\pi/3$ momenta. Taking the imaginary part, we get that 

\begin{widetext}
\begin{align}
&\Imp[ 2G_0 G_{\pm} + G_{\pm}^2] = 2\Imp G_0 \Rep G_{\pm} + 2\Rep G_0 \Imp G_{\pm} + \Imp G_{\pm}^2 \notag \\
&=\lim_{\eta \rightarrow 0} \left( -2\pi \delta(\omega - \lambda - \Delta) \frac{\omega - \lambda + 2\Delta}{(\omega - \lambda + 2\Delta)^2 + \eta^2}  -2\pi \delta(\omega - \lambda +2 \Delta) \frac{\omega - \lambda - \Delta}{(\omega - \lambda - \Delta)^2 + \eta^2}   + 2\pi \delta'(\omega - \lambda - \Delta) \right) \; ,
\end{align}
\end{widetext}

\noindent The integral over frequencies weighted by the Bose function for the third term gives the usual $2\beta n_B (\lambda + \Delta) (1 + n_B (\lambda + \Delta))/3$. As for the first two terms, taking the limit of small $\eta$ and performing the integral, we get $2(n_B(\lambda + \Delta) - n_B (\lambda - 2\Delta))/9\Delta$. Putting these two together, we get 

\begin{equation}
\begin{split}
\chi_{\pm} &= \frac{2\beta}{3} n_B (\lambda + \Delta) (1 + n_B (\lambda + \Delta)) \\
& \; \; \;+ \frac{2}{9\Delta}(n_B (\lambda - 2\Delta) - n_B(\lambda + \Delta)) \; .
\end{split}
\end{equation}

For $T \gg \Delta$, we again use that $n_B (\lambda + \Delta) = n_B (\lambda  - 2\Delta) = n_B (\lambda)$, as well as the high-temperature limit of $\lambda$ so that $ n_B (\lambda^\text{High T}) = {q}$. This leads to $\chi_{\pm} \rightarrow {\frac{2q (1 + q)}{3T}}$. For the other limit where $\Delta \gg T$ while still being at high-temperatures, we approximate {$n_B (\lambda + \Delta) = 0$ and $n_B (\lambda  - 2\Delta) = 3 q$. This leads to $\chi_{\pm} \rightarrow  \frac{2q}{\Delta}$}, which produces one of the results in table~\ref{tableA} in the main text for the intersite susceptibility.

Finally, since $\chi_{loc} = \frac{1}{3}[\chi_0 + 2 \chi_{\pm}]$, it will share the same high-temperature limit as $\chi_{0}$ and $\chi_{\pm}$ at $T \gg \Delta$ and be Curie-like. At intermediate temperature, i.e. when $\Delta \gg T$, then the $\chi_{\pm}$ contribution is outshone since it does not increase with decreasing temperature. The behavior of $\chi_{loc}$ will then be non-trivial, but nevertheless we can ascertain that $\chi_{loc} < \chi_0$, and will not present the same clear intermediate plateau behavior. 

Note that the interpretation of these limits is readily extended to the case of $J_H$ vs $T$, as $\Delta = {\rm O}(1) J_H$ at most. The different limiting cases are succintly presented in table~\ref{tableA} in the main text, using the general functions

\begin{align}
{\tilde S} (x) &= (1+x) \ln{(1+x)} - x \ln{x}, \\
{\tilde\chi} (x,T) &= 2x(1+x)/T .
\end{align}

This concludes the derivation of the limits presented in table~\ref{tableA}, as well as in figure~\ref{fig:figure_4}.

\section{Details for single iteration approach} \label{appendix2}

In this appendix, we proceed to derive the first part of $\Rep \Sigma_{\chi} (\omega = 0 + i \eta)$ (see Eq.~\ref{eq:conditionSI}). Knowing that the imaginary part of the bare local bosonic Green's function is

\begin{equation}
G_{B, \rm loc}'' (\omega) = -\pi \left( \frac{1}{3} \delta(\omega - \lambda') +  \frac{2}{3} \delta(\omega - \lambda'- 3\Delta)\right),
\end{equation}

\noindent then, together with the real part of $g_c$ from equation~\eqref{eq:elG}, results in

\begin{align}
\Rep \int &\frac{d \omega}{\pi} G_{B, \rm loc}'' (\omega) n_B(\omega) g_c(\omega) \notag \\
&= \frac{2\rho}{3} n_B(\lambda' + 3\: \Delta) \ln \left( \frac{|\lambda' + 3 \: \Delta -D|}{|\lambda' + 3\: \Delta + D|} \right) \notag \\  
& + \frac{\rho}{3} n_B(\lambda') \ln \left( \frac{| \lambda' -D|}{|\lambda' + D|} \right)  \rightarrow 0 \;.
\label{eq:part1neg}
\end{align}

For the physical regimes here considered, $\lambda' \propto T \ll \min \{ J_H, D\}$, we have that for the term proportional to $n_B(\lambda')$, since the accompanying logarithm tends to $0$ due to the cancellation of $D$. The leftover term is also $0$ in the limit where $D \gg J_H$ due to the logarithm. This logarithm factor is however finite for $\Delta \propto J_H \gg D$. In this case, the solution of Eq.~\eqref{eqs:qDelta2} shows that $n_B (\lambda' + 3 \Delta) \rightarrow 0$. Hence this complete term of Eq.~\eqref{eq:part1neg} can be correctly neglected compared to the other contribution, as is mentioned in section~\ref{subsecTKsi}.

Note that this result was obtained with a symmetric electron dispersion. In the case of particle-hole asymmetry of the conduction electrons, this integral would not tend to $0$, but rather to a finite number which has the effect of changing the effective value of $J_K$. This is expected as the particle-hole asymmetry creates potential scattering that contributes to the renormalization of the Kondo coupling~\cite{hewson1997kondo}.


\bibliography{biblio}

\end{document}